\documentclass[a4paper,11pt]{article}
\usepackage{html,amsmath,amssymb,amsfonts,epsfig,cancel,cite,longtable,caption,siunitx,array,color,booktabs,setspace}

\newlength{\xtrawidth}
\newlength{\xtraheight}
\newcommand{\be}{\begin{equation}}
\newcommand{\ee}{\end{equation}}
\newcommand{\beq}{\begin{equation}}
\newcommand{\eeq}{\end{equation}}
\newcommand{\ba}{\begin{array}}
\newcommand{\ea}{\end{array}}
\newcommand{\bea}{\begin{eqnarray}}
\newcommand{\eea}{\end{eqnarray}}
\newcommand{\bean}{\begin{eqnarray*}}
\newcommand{\eean}{\end{eqnarray*}}
\newcommand{\eref}[1]{(\ref{#1})}

\newcommand{\IP}{\mathbb{P}}

\newcommand{\IZ}{\mathbb{Z}}
\newcommand{\cO}{{\cal O}}
\newcommand{\cN}{{\cal N}}

\def\fnote#1#2{\begingroup\def\thefootnote{#1}\footnote{#2}
     \addtocounter{footnote}{-1}\endgroup}

\newcommand{\cicy}[2]{\begin{matrix} #1\end{matrix}\!\left[\begin{matrix}#2 \end{matrix}\right]}

\newcommand*{\myalign}[2]{\multicolumn{1}{#1}{#2}}
\newcommand{\varstr}[2]{\vrule height #1 depth #2 width0pt}

\numberwithin{equation}{section}

\begin{document}

\vspace{1cm}

\title{{\Large \bf A Comprehensive Scan for Heterotic $SU(5)$ GUT models}}

\vspace{2cm}

\author{
Lara B. Anderson${}^{1}$,
Andrei Constantin${}^{2}$,
James Gray${}^{3}$,\\[4pt]
Andre Lukas${}^{2}$ and
Eran Palti${}^{4,5}$\\[-0.5cm]
}

\date{}
\maketitle
{\setstretch{1.1}
\begin{center} {\small ${}^1${\it Center for the Fundamental Laws of Nature,  \\
Jefferson Laboratory, Harvard University, \\ 17 Oxford Street, Cambridge, MA 02138, U.S.A.\\[0.3cm]
       ${}^2$Rudolf Peierls Centre for Theoretical Physics, Oxford
       University,\\
       $~~~~~$ 1 Keble Road, Oxford, OX1 3NP, U.K.\\[0.3cm]
      ${}^3$Arnold-Sommerfeld-Center for Theoretical Physics, \\
       Department f\"ur Physik, Ludwig-Maximilians-Universit\"at M\"unchen,\\
       Theresienstra\ss e 37, 80333 M\"unchen, Germany\\[0.3cm]
       ${}^4$Centre de Physique Theorique, Ecole Polytechnique, CNRS, 91128 Palaiseau, France.\\[0.3cm]
       ${}^5$Institut f\"ur Theoretische Physik, Universitat Heidelberg, \\
       Philosophenweg 19, D-69120 Heidelberg, Germany}}\\

\fnote{}{lara@physics.harvard.edu, a.constantin1@physics.ox.ac.uk, james.gray@physik.uni-muenchen.de}
\fnote{}{lukas@physics.ox.ac.uk, Eran.Palti@cpht.polytechnique.fr} 
\end{center}
}

\begin{abstract}
\noindent Compactifications of heterotic theories on smooth Calabi-Yau manifolds remains one of the most promising approaches to string phenomenology. In two previous papers,  \htmladdnormallink{{\tt arXiv:1106.4804}}{http://arXiv.org/abs/arXiv:1106.4804} and \htmladdnormallink{{\tt arXiv:1202.1757}}{http://arXiv.org/abs/arXiv:1202.1757}, large classes of such vacua were constructed, using sums of line bundles over complete intersection Calabi-Yau manifolds in products of projective spaces that admit smooth quotients by finite groups. A total of $10^{12}$ different vector bundles  were investigated which led to $202$ $SU(5)$ Grand Unified Theory (GUT) models. With the addition of Wilson lines, these in turn led, by a conservative counting, to $2122$ heterotic standard models. In the present paper, we extend the scope of this programme and perform an exhaustive scan over the same class of models. A total of $10^{40}$ vector bundles are analysed leading to $35,000$ $SU(5)$ GUT models. All of these compactifications have the right field content to induce low-energy models with the matter spectrum of the supersymmetric standard model, with no exotics of any kind. The detailed analysis of the resulting vast number of heterotic standard models is a substantial and ongoing task in computational algebraic geometry.
\end{abstract}

\thispagestyle{empty}
\setcounter{page}{0}

\newpage

{\setstretch{1.7}
\thispagestyle{empty}
\tableofcontents
}

\newpage
%

\section{Introduction and Summary}

Heterotic string compactifications~\cite{Gross:1984dd, Gross:1985fr, Gross:1985rr, Candelas:1985en} on Calabi-Yau threefolds have provided one of the most promising approaches to string phenomenology for almost three decades. Several approaches have been proposed and used over the years: smooth Calabi-Yau compactifications based on the standard embedding \cite{Candelas:1985en, Greene:1986bm, Greene:1986jb, Braun:2009qy, Braun:2011ni}, non-standard embedding models \cite{Distler:1987ee, Distler:1993mk, Kachru:1995em, Braun:2005ux, Braun:2005bw, Braun:2005nv, Bouchard:2005ag, Blumenhagen:2006ux, Blumenhagen:2006wj, Anderson:2007nc, Anderson:2008uw, Anderson:2009mh}, models based on orbifolds \cite{Buchmuller:2005jr, Buchmuller:2006ik, Lebedev:2006kn, Lebedev:2007hv, Lebedev:2008un, Kim:2007mt, Nibbelink:2009sp, Blaszczyk:2009in, Blaszczyk:2010db, Kappl:2010yu}, free fermionic strings \cite{Assel:2009xa, Christodoulides:2011zs, Cleaver:2011ir} and Gepner models \cite{GatoRivera:2009yt, GatoRivera:2010xn, Maio:2011qn}. 
In the present paper, we display the latest results of a large scale model building programme in the context of smooth Calabi-Yau compactifications of the heterotic string. This programme, aimed at achieving more detailed phenomenology than has to date be possible in this context, was initiated in the publications \cite{Anderson:2011ns, Anderson:2012yf}. The history of string phenomenology suggests that it is difficult to fine tune any particular construction in order to simultaneously meet all the properties of the Standard Model. 
Instead, the approach we take is that of a `blind' automated scan over a huge number of models; for the present scan this number is of order $10^{40}$.  

Following this approach, what lies in front of the heterotic string model builder is  a set of highly non-trivial challenges that can be summarised in the following checklist:
\begin{itemize}
	\item[1.] Construct a geometrical set-up, such that the 4-dimensional compactification of the $\cN=1$ supergravity limit of the heterotic string contains the symmetry $SU(3)\times SU(2)\times U(1)$ of the Standard Model of particle physics. This step is usually realised in two stages, by firstly breaking the $E_8$ heterotic symmetry to a Grand Unified Theory (GUT) group and then breaking the latter to the Standard Model gauge group (plus possibly $U(1)$ factors).  This requires a VEV of the gauge connection on the internal (compact) 6-dimensional space $X$, or, equivalently, one needs to construct a vector bundle $V\rightarrow X$. 
	\item[2.] Derive the matter spectrum of the 4-dimensional theory. At low energy, the fermion fields transforming under the broken gauge group must be massless modes of the Dirac operator on the internal space $X$. The number of massless modes for a given representation is given by the dimension of certain bundle-valued cohomology groups on $X$. Such cohomology computations are generically difficult to perform. At this stage, one would like to identify models with the matter spectrum of the minimally supersymmetric Standard Model (MSSM), 
typically a very small sub-set of all consistent models constructed in this way.
	\item[3.] Constrain the resulting Lagrangian, in order to avoid well-known problems of supersymmetric GUT models, such as fast proton decay. For this purpose, additional discrete or continuous symmetries derived from the compactification set-up may be helpful.
	\item[4.] Derive information about the detailed properties of the model, such as the superpotential, the holomorphic Yukawa couplings, fermion mass-terms and $\mu$-terms. Such holomorphic quantities can usually be understood using techniques from algebraic geometry.
	\item[5.] Compute the physical Yukawa couplings. The physical Yukawa couplings consist of holomorphic superpotential terms times a non-holomorphic prefactor, whose computation requires the explicit knowledge of the metric on $X$ and the gauge connection on the vector bundle $V$. For the case when $X$ is a Calabi-Yau manifold, Yau's proof \cite{Yau:1978} guarantees the existence of a Ricci-flat metric, while for poly-stable vector bundles on Calabi-Yau manifolds, the Donaldson-Uhlenbeck-Yau theorem \cite{Donaldson1985, Uhlenbeck1986} guarantees the existence of a Hermitian Yang-Mills connection. However, except in very special cases, these quantities are not known analytically. So far, one can approach this differential geometric problem only numerically \cite{Douglas:2006hz,Anderson:2010ke,Anderson:2011ed}. 
	\item[6.] Stabilize the moduli and break supersymmetry. Recently, some progress has been made by including the effect of the $E_8\times E_8$ bundle flux~\cite{Anderson:2011cza, Anderson:2011ty, Anderson:2013qca}. 
	\item[7.] Compute soft-breaking parameters.
\end{itemize}
Every phenomenological requirement in this list will lead to a substantial reduction in the number of viable models. It is, therefore, crucial to start with a large number of models at the initial stages, if one hopes to retain a realistic model in the end. In this paper, we will concentrate on precisely this task and obtain, within a certain class of constructions, the largest possible set of models after the first two steps.

The history of this field can be largely understood by looking at the types of poly-stable holomorphic vector bundles that have been the focus of study at any given time. In the early days of the subject researchers largely concentrated on small deviations from the ``standard embedding", where the gauge bundle was taken to be a holomorphic deformation of the tangent bundle \cite{Greene:1986bm, Greene:1986jb}. Such work has been continued to the current day with the first exact MSSM being produced from such an approach relatively recently \cite{Braun:2011ni}. In the 1990's and later more general poly-stable holomorphic vector bundles, or ``non-standard embeddings", began to be considered in ernest \cite{Distler:1987ee, Distler:1993mk, Kachru:1995em, Braun:2005ux, Braun:2005bw, Braun:2005nv, Bouchard:2005ag, Blumenhagen:2006ux, Blumenhagen:2006wj, Anderson:2007nc, Anderson:2008uw, Anderson:2009mh}. These gauge bundles were typically taken to have structure groups $SU(3)$, $SU(4)$ or $SU(5)$ leading to an $E_6$, $SO(10)$ or $SU(5)$ GUT group, respectively.

Recently, a new approach to building heterotic models on smooth Calabi-Yau threefolds has been advanced~\cite{Anderson:2011ns,Anderson:2012yf}. In this approach, the vector bundles in consideration were chosen to be simple sums of line bundles. This construction leads to a GUT group which naively includes additional $U(1)$ factors in the GUT group beyond the gauge groups mentioned above. However, these extra $U(1)$'s are frequently broken, in addition to other effects, by the Green-Schwarz mechanism. As such, these models are just as capable of leading to acceptable particle physics phenomenology as their non-abelian cousins.

There are several advantages to working with sums of line bundles, as opposed to irreducible vector bundles. Firstly, such configurations are relatively simple to deal with from a computational point of view, and as a result, vastly greater numbers of models can be considered as compared to other approaches, such as \cite{Anderson:2007nc, Anderson:2008uw, Anderson:2009mh}. Secondly, although broken at a high scale, the additional $U(1)$ symmetries that are present in these models can greatly constrain the Lagrangian of these models giving more information about the superpotential, and in particular the K\"ahler potential, than is usually available in smooth heterotic constructions.  Finally, although line bundle sums often represent special loci in the moduli space of vector bundles of a given topology, one can move away from the `split' locus by turning on VEVs for certain bundle moduli, thus reaching non-abelian bundles. As such, these simple configurations provide a computationally accessible window into an even bigger moduli space of heterotic compactifications.

The two previous papers in which the line bundle construction was employed, \cite{Anderson:2011ns,Anderson:2012yf}, achieved a number of goals. These publications presented the results of a scan over some $10^{12}$ line bundle sums in a search for heterotic standard models. The number of models investigated was in some sense arbitrary. The range of integer values defining the first Chern class of the line bundles scanned over was relatively restricted. These values were chosen simply to give a large number of models which was, nevertheless, manageable with a relatively simply implemented algorithm. From these $10^{12}$ vector bundles, the authors extracted $202$ $SU(5)$ GUT models which had precisely three generations of GUT families, no anti-families, at least one $5-\overline{5}$ pair of Higgs fields and no other charged matter of any kind whatsoever. Each of these models was constructed such that there was at least one Wilson line which could be added to the configuration which lead to exactly the spectrum of the MSSM with the Higgs triplets being projected out. In fact, each of these GUT models lead to many different standard models due to choices, closely related to the Wilson line breaking, which were available in the construction. The results of including Wilson lines were presented explicitly and the $202$ GUT models led to, by a conservative counting, $2122$ heterotic standard models. The constraints on the effective field theory description of these models coming from the broken $U(1)$ gauge factors was also explicitly computed. The complete data set including higher dimensional construction and tabulation of the resulting four-dimensional effective field theories was presented in a data base which can be found here \cite{database}.

In this paper, we extend the scope of the scan to make it comprehensive within the class of heterotic compactifications being studied. Instead of restricting the range of integers defining the line bundles in an arbitrary manner, we developed an algorithm which allows for an exhaustive scan. This leads to a considerable increase in the size of the data set being considered. Instead of examining $10^{12}$ configurations as before, in this paper we present the results of a scan over $10^{40}$ different compactifications of heterotic string theory. Even with improvements to our methodology from the previously published work, this scan, which is described in Section \ref{sec:alg} ran on a computer cluster over a period of seven months. The data set of GUT models we have obtained is, unsurprisingly, much larger than the $202$ models discussed above. A conservative counting results in $34,989$ such GUT models which we expect will lead to one order of magnitude more heterotic standard models when Wilson lines are added. Given the size and extra technical complications resulting from dealing with such huge numbers of heterotic compactifications, we will present the results of our investigations in two separate publications. In particular, the detailed analysis of  incorporating the effects of the Wilson line breaking, will be presented in a separate paper. This represents in and of itself a huge task in computational algebraic geometry, which will take several months to complete.

The table below presents a statistics on the total number of consistent GUT models which have resulted from the search detailed in Section \ref{sec:alg}. The first column counts $SU(5)$ GUT models having the correct chiral asymmetry, which can, however, suffer from the presence of $\mathbf{\overline{10}}$ multiplets or the absence of $\mathbf{5}-\overline{\mathbf{5}}$ pairs 
 to accomodate the Higgs content of the standard model. In the second column we eliminate those models that contain $\mathbf{\overline{10}}$ anti-family matter. This step relies on computations of line bundle cohomology groups, which we are able to perform in 94\% of all cases. The number in parentheses indicates the GUT models for which we could not decide upon the presence of $\mathbf{\overline{10}}$ multiplets. Similarly, in the third column we select from the $44343$ models that definitely have no anti-families, those which contain at least one $\mathbf{5}-\overline{\mathbf{5}}$ pair to contain MSSM Higgs fields.

\vspace{10pt}
\begin{footnotesize}
\begin{longtable}{| c | c | c |c|}
\captionsetup{width=14cm}
\caption{\label{basicstat}\it  Statistics on the number of models:}\\
\hline
\myalign{| m{2.2cm}|}{$\ $GUT models} &
\myalign{m{3.5cm}|}{ $\ \ \ \ $ no $ \overline{\mathbf{10}}$ multiplets$\ \ \ $ }&
\myalign{m{3.5cm}|}{$\ \ \ \ \ \ \ $ no $ \overline{\mathbf{10}}\,$s  and $\ \ \ \ \ \ \ $ at least one $\mathbf{5}-\overline{\mathbf{5}}$ pair}
\\ \hline
63325 & 44343 (3606) & 34989 (5291) 
\\ \hline
\end{longtable}
\end{footnotesize}
\vspace{10pt}

The rest of this paper is structured as follows. In the next section we recapitulate the line bundle construction, emphasising the discussion of the structure group of holomorphic sums of line bundles, which is crucial in determining correctly the GUT gauge group. The discussion is new and, for the purpose of preserving the fluidity of the text, we defer a full presentation to Appendix \ref{Struc_Group_App}. In the following two sections we define the class of manifolds under consideration and present the constraints imposed on the vector bundles. In Section~\ref{sec:alg} we outline the algorithm used in the automated scan, while in Section~\ref{sec:results} we list the number of viable models obtained over each manifold, noting that in all cases we reach a limit beyond which no realistic line bundle vacua exist. We conclude with an example and final remarks.

\section{Overview of the Construction}
The structure of $E_8\times E_8$ heterotic compactifications on smooth Calabi-Yau three-folds with Abelian vector bundles, as well as the class of $\mathcal N=1$ four dimensional supergravities to which they lead, have been thoroughly discussed in two previous publications \cite{Anderson:2011ns, Anderson:2012yf}. As such, we limit the scope of this section to merely summarising the central features of heterotic line bundle standard models. Additionally, we provide a discussion on the possible structure groups of vector bundles constructed as direct sums of holomorphic line bundles.

\subsection{Heterotic Line Bundle Compactifications}

Schematically, the construction and analysis of heterotic string line bundle standard models can be broken up into three steps.
 \begin{itemize}
\item[$1.$]  In the first step, a solution to the 10-dimensional supergravity limit of the $E_8\times E_8$ heterotic string is obtained by specifying several geometrical elements. Firstly, we compactify 10-dimensional space-time on a smooth Calabi-Yau threefold $X$. Over this manifold we specify a poly-stable holomorphic vector bundle $V$ with structure group $H \subset E_8 \times E_8$ which describes the gauge field expectation values in the supergravity solution. The possible choices of $V$ over a given $X$ are restricted by several consistency requirements as described in Sections \ref{gut_group} and \ref{bundle_sec}. In the line bundle construction, the vector bundle is taken to be a direct sum of five holomorphic line bundles  $$V = \bigoplus_a L_a\; .$$ As we will discuss in Section~\ref{gut_group}, we choose the five line bundles $L_a$ such that the structure group $H \subset E_8$ is Abelian and of the form $H =S\left(U(1)^5\right)\cong U(1)^4$. 

If the background derived in this first step was used to dimensionally reduce the heterotic string theory to obtain an $\cN=1$ four dimensional supergravity without further modification, then the result would be a supersymmetric GUT. The gauge group seen in four dimensions $G$ would, naively, be the commutant of $H$ inside $E_8$. For the line bundle models mentioned above, this leads to a GUT group $$G = SU(5)\times S\left(U(1)^5\right)$$ However, the additional $U(1)$ factors are generically Green-Schwarz anomalous and thus the associated gauge bosons often obtain St\"uckelberg masses which are close to the compactification scale in magnitude. 

\item[$2.$] In the second step, Wilson lines are added on the Calabi-Yau in such a way as to break the GUT group described above, down to that of the Standard Model. Adding such structure to the compactification is only possible if $X$ is not simply connected. Most standard constructions of Calabi-Yau threefolds lead to manifolds for which $\pi_1(X)=0$. Fortunately this situation can be resolved by quotienting a monifold $X$ obtained from one of the usual constructions by a freely acting discrete symmetry $\Gamma$. The fundamental group of the resulting smooth quotient manifold $\hat{X}=X/\Gamma$ is non-trivial, and in fact is isomorphic to $\Gamma$. 

The vector bundle $V$ constructed in step 1 must be consistent with this quotienting procedure. We must ensure that our bundle $V\rightarrow X$ descends to a well defined vector bundle $\hat{V} \rightarrow \hat{X}$. This is only the case if $V$ admits an equivariant structure under the symmetry $\Gamma$. Indeed, the set of vector bundles on $\hat{X}$ is in one-to one correspondence with the set of equivariant vector bundles on $X$.

The heterotic theory is then compactified to four dimensions on this new quotiented configuration including a non-trivial Wilson line. The gauge group obtained in four dimensions is then the commutant of the structure group of the flat bundle associated to the Wilson line inside $G$. This result is corrected as described in the first step by the Green-Schwarz mechanism. If the configuration is chosen correctly this can lead to the standard model gauge group $G_{SM}$ in four dimensions. The matter content must be computed by the usual techniques of dimensional reduction - including the effects of the Wilson line. One wishes to obtain examples where the resulting four dimensional standard model charged matter is exactly that of the MSSM.

\item[$3.$] As a final step in analysing a heterotic line bundle standard model, one can use global remnants of the additional $U(1)$ four dimensional gauge symmetries which are broken by the Green-Schwarz mechanism to constrain the operators present in the four dimensional Lagrangian. This allows a degree of analytical control over the low energy theory associated to these models which is unusual in the context of smooth Calabi-Yau reductions - in particular with regards to the K\"ahler potential for matter fields. In specific models, these symmetries can forbid operators in the four dimensional theory whose presence can be problematic for issues such as proton stability. 

\end{itemize}

In this paper we present the results obtained after pursuing the avenue described in the first step described above. We construct a large class of GUT models, postponing the remaining analysis for a future publication. However, we stress that the full analysis is feasible and has already been carried out for the more restricted set of models described in Ref.~\cite{Anderson:2011ns, Anderson:2012yf}. In the rest of this section we describe the GUT gauge group and particle spectrum that is obtained in such constructions in more detail.

\subsection{The GUT Gauge Group}\label{gut_group}
As discussed in Refs.~\cite{Anderson:2011ns, Anderson:2012yf}, the gauge group of the GUT models that we hope to construct using line bundles is $SU(5)\times S\left(U(1)^5\right)$, the maximal subgroup of $E_8$ which commutes with $S\left( U(1)^5\right)$. We reserve this section for determining when the structure group of a direct sum of five holomorphic line bundles with vanishing first Chern class is indeed $S\left( U(1)^5\right)$, leading to the desired GUT  gauge group. We will outline below the possible structure groups (and obstructions) for $V=\bigoplus_{a=1}^{5} L_a$.

It is a long standing problem in vector bundle geometry that in general, the structure group, $H$, of a vector bundle cannot be determined without explicit knowledge of the $H$-valued connection (satisfying the equations of motion, here the Hermitian Yang-Mills equations \cite{Anderson:2011ns}). However, for holomorphic bundles in certain cases, knowledge of the topology of the bundle and other facts may be enough to fully specify $H$.

As discussed more fully in Appendix \ref{Struc_Group_App}, for the present scans of rank $5$, reducible bundles, built as a sum of holomorphic line bundles $\bigoplus_{a=1}^{5} L_{a}$, there are only a few possibilities for $H$. We demand that $\sum_{a} c_1(L_1)=0$. Thus, it can be argued $H$ must be a subgroup of $SU(5)$, $SO(5)$ or $Sp(4)$ (see Appendix \ref{Struc_Group_App}). The latter two structure groups are possible for a rank $5$ vector bundle only if $V_5$ admits either a real or symplectic fiber structure (see \cite{huybrechts2005complex,zbMATH01538887} and \eref{fiber_struc}), in the form of a vector bundle isomorphism, $\phi: V \to V^{*}$. Since $\bigoplus_a L_a$ is an odd sum of $5$ line bundles, such an isomorphism is possible if and only if $L_a =\cO_{X}$ for at least one $a$. To avoid this case we impose for all $a=1,\ldots ,5$ that 
\beq\label{c_not_triv_first}
c_1^r(L_a) \neq 0 \quad\text{for at least one value of}\quad r=1,\ldots h^{1,1}(X)\; .
\eeq
This constraint, combined with the vanishing of the first Chern class, means that $H$ must be a sub-group of $SU(5)$. It only remains to determine whether $H=S\left(U(1)^{5}\right)$ or a proper sub-group thereof. If $H$ is not equal $S\left(U(1)^{5}\right)$, it is possible that its commutant in $E_8$ (the $4d$ GUT group, $G$) is not of the form $SU(5)\times U(1)^4$, but rather another group less suitable for realistic model-building. For example, the structure group $H=S\left(U(1)^2\right) \times S\left(U(1)^3 \right)$ has commutant, $G=SU(6)\times U(1)^3$. To eliminate such phenomenologically unviable possibilities, the following condition is imposed on the Chern classes of the line bundle sum:

\beq\label{no_subgroup_first}
\sum_{a\in S} c_1(L_{a})\neq 0\;\; \text{for all proper subsets}\;\;  S\subset\{1,\ldots ,5\}\; .
\eeq

With these conditions in hand, it is guaranteed that heterotic line bundle construction leads to a $4d$ GUT symmetry of the form $SU(5)\times U(1)^{4}$ (with the abelian factors generically Green-Schwarz massive). We are now ready to turn to the more detailed question of the charged matter particle spectrum of the low-energy theory.

\subsection{The GUT spectrum} \label{spec}

If the conditions of the previous section are satisfied, a sum of five line bundles breaks the $E_8$ heterotic symmetry to $SU(5)\times S\left(U(1)^5\right)$. In such a case, the computation of the spectrum of the heterotic line bundle model has been explained in detail in \cite{Anderson:2011ns,Anderson:2012yf}. Here we simply state the results for the convenience of the reader.

We represent $S\left( U(1)^5\right)$ representations by vectors $\mathbf{q} = \left( q_1, \ldots, q_5\right)$ of five integer charges. Due to the determinant condition, two such vectors ${\bf q}$ and ${\bf q}'$, represent the same representation and, hence, have to be identified if $\mathbf{q}-\mathbf{q'} \in \IZ\mathbf{n}$, where ${\bf n}=(1,1,1,1,1)$. We also introduce the standard basis $\{\mathbf{e}_a\}_{a=1,\ldots,5}$ in five dimensions. Charges for GUT multiplets are indicated by adding the charge vector as a subscript so that, for example, ${\bf 10}_{{\bf e}_1}$ represents a ${\bf 10}$ multiplet of $SU(5)$ with charge 1 under the first $U(1)$ and uncharged under the others. A list of the relevant multiplets and their properties is provided in Table~\ref{spectrum}, below.
\begin{table}[!h]
\begin{center}
\begin{tabular}{|l|l|l|l|l|l|l|l|}
\hline
\varstr{14pt}{9pt} repr. & cohomology & total number & required for MSSM \\ \hline\hline
\varstr{14pt}{9pt} ${\bf 1}_{{\bf e}_a - {\bf e}_b}$ & $H^1(L_a \otimes L_b^*)$  &  $\sum_{a,b} h^1(L_a \otimes L_b^*) = h^1(V \otimes V^*)$ & \;\;\;\;\; - \\ \hline
\varstr{14pt}{9pt} ${\bf 5}_{-{\bf e}_a -{\bf e}_b}$ & $H^1(L_a^* \otimes L_b^*)$  & $\sum_{a<b} h^1(L_a^* \otimes L_b^*) =h^1(\wedge^2 V^*) = h^1(\wedge^2 V) $ & \;\;\;\;\;$n_h$\\ \hline
\varstr{14pt}{9pt} ${\bf \overline{5}}_{{\bf e}_a+{\bf e}_b}$ & $H^1(L_a \otimes L_b)$  & $\sum_{a<b} h^1(L_a \otimes L_b) =h^1(\wedge^2 V) $ & \;\;\;\;\;$3 |\Gamma| + n_h$\\ \hline
\varstr{14pt}{9pt} ${\bf 10}_{{\bf e}_a}$ &$H^1(L_a)$ & $\sum_a h^1 (L_a) = h^1 (V)$& \;\;\;\;\;$3 | \Gamma|$\\ \hline
\varstr{14pt}{9pt} ${\bf  \overline{10}}_{-{\bf e}_a}$ & $H^1(L_a^*)$ & $\sum_a h^1(L_a^*) = h^1(V^*)$&\;\;\;\;\; 0
\\ \hline 
 \end{tabular}
 \vskip 0.4cm
\parbox{16.7cm}{\caption{\it\small The possible $SU(5)$ matter representations which may be obtained in four dimensions and their $U(1)$ charges. The dimensions of the cohomology groups indicated in the second column determines the multiplicity of each representation in the four dimensional spectrum. The third column gives the total number of each $SU(5)$ representation present in the four dimensional effective theory, of any $U(1)$ charge. The final column gives the number of each $SU(5)$ multiplet that we require in the GUT theory in order to obtain the standard model spectrum (with $n_h$ pairs of Higgs doublets) after the addition of suitable Wilson lines.}\label{spectrum}}
 \end{center}
 \end{table}
 Different $SU(5)$ representations are associated with different patterns of $U(1)$ charges. For example, the ${\bf 10}$ multiplets carry charge one under precisely one of the five $U(1)$ symmetries, while the ${\bf \overline{5}}$ multiplets carry charge one with respect to two $U(1)$ symmetries. Apart from such rules, the precise assignment of charges across the spectrum (including bundle moduli) is model-dependent. This is of major phenomenological importance: invariance under the (global remnant of the) $S\left( U(1)^5\right)$ symmetry constrains the allowed operators in the low-energy theory. Indeed, one can easily envisage situations in which the pattern of charges is such that, for example proton decay operators are forbidden. 
 
The final column of the table shows the number of each $SU(5)$ representation that we require in the GUT model such that after quotienting the Calabi-Yau manifold and adding suitable Wilson lines, we obtain the spectrum of the MSSM. As before, $\Gamma$ denotes a freely acting finite group by which we quotient the Calabi-Yau manifold and $|\Gamma|$ is its order.
 
We have now completed our overview of what is required for a successful heterotic line bundle standard model construction. In the next sections we move on to describe the particular class of Calabi-Yau threefolds which we will study, as well as some details of the line bundles over them.

\section{The Manifolds}
\label{Sec3}
Historically the first class of Calabi-Yau three-folds explicitly constructed \cite{Candelas:1987kf}, complete intersections in products of projective spaces (CICYs) have often served as the starting point in heterotic model building \cite{Greene:1986jb, Greene:1986bm, Braun:2009qy, Braun:2005ux, Braun:2005bw, Braun:2005nv, Bouchard:2005ag, Anderson:2007nc, Anderson:2008uw, Anderson:2009mh, Braun:2011ni}. The present systematic computer-based scan for standard models will continue this tradition.

The choice is based on two crucial features of the CICY class of manifolds. In the first place, there exists a systematic classification of all linearly realised freely acting discrete symmetries on the CICY manifolds in the database \cite{Candelas:2008wb, Braun:2010vc}. More accurately, Braun's classification \cite{Braun:2010vc} provides a list of all such symmetries which descend from a linearly acting symmetry on the ambient space. Furthermore, a given Calabi-Yau manifold can frequently be embedded in many different products of projective spaces. The symmetry classification of \cite{Braun:2010vc} is carried out for a limited selection of possible ambient spaces for each Calabi-Yau. As discussed in the previous section, a knowledge of such symmetries is an essential ingredient in breaking the GUT group to the Standard Model gauge group. We note that in constructing the GUT models presented here, it is in fact only knowledge of the possible orders of the available groups, and thus the possible values of $|\Gamma|$, which is required.

The second feature of the CICY's which makes them particularly suitable for the current work is related to their relative simplicity. The embedding of Calabi-Yau manifolds in such simple ambient spaces means that computations of the cohomology of line bundles over these manifolds can be effectively automatised on a computer \cite{cicypackage}.    
This is particularly true when there is a strong connection between line bundles on the ambient space and line bundles on the Calabi-Yau threefold. As such we restrict our attention to CICYs presented in a `favourable' embedding.

Favourable embeddings can be described in many equivalent ways. For example, on manifolds such as those we are considering, isomorphism classes of line bundles are completely classified by their first Chern class. This is an element of the second cohomology group $H^2(X,\IZ)$ of the base space $X$. Favourable CICYs can be defined to be those whose second cohomology descends entirely from the second cohomology of the embedding space. In such a case, all of the line bundles on $X$ are restrictions of line bundles over the ambient product of projective spaces. For this property to hold for a given description of a CICY, certain requirements have to be satisfied, as discussed in Appendix \ref{appB}.

\section{The Bundles}\label{bundle_sec}
Recall that our bundles $V\rightarrow X$ are taken to be sums of five line bundles 
\begin{equation}\label{Vdef}
\displaystyle V=\bigoplus_{a=1}^{5} L_a\; .
\end{equation}
A single line bundle is specified by its first Chern class and, hence, by a set of $h^{1,1}(X)$  integers. Given this, a sum of $5$ line bundles is specified by a matrix of integers with $h^{1,1}(X)$ rows and $5$ columns. In our systematic investigation, we have scanned over $\sim\!\!10^{40}$ such matrices and selected approximately $35,000$ bundles which lead to phenomenologically consistent $SU(5)$ GUTs. In the following subsections we list the criteria that these $35,000$ models satisfy.

\subsection{Topological Constraints}
We require that,
\begin{eqnarray} \label{C1}
c_1(V) = 0 \;,  
\end{eqnarray}
as well we Eq.~\eqref{no_subgroup_first}, such that the structure group of $V$ is $S(U(1)^5)$ which leads to a GUT group $G = SU(5)\times S\left(U(1)^5\right)$. Apart from this group theoretical advantage, imposing~$c_1(V)=0$ guarantees the existence of a spin structure on $V$.

In addition, the integrability condition on the Bianchi Identity for the Neveu-Schwarz two form leads to the following constraint on the vector bundle $V$.
\beq\label{anom_canc}
{\rm ch}_2(TX)-{\rm ch}_2(V) -{\rm ch}_2(\tilde V)= [C]
\eeq
Here $[C]\in H^4(X)$ is the Poincar\'e dual to the effective holomorphic curve class  wrapped by a five-brane and~$\tilde V$ is the hidden-sector bundle (which we will take to be trivial). The simplest way to guarantee that this condition can be satisfied is to require that
$c_2(TX)-c_2(V)\in $ Mori cone of $X$, where we have used that $c_1(V)=0$. In this case, an effective curve class which saturates the condition~\eqref{anom_canc} for a trivial hidden bundle $\tilde{V}$ exists (although, typically, solutions with non-trivial $\tilde{V}$ can be found as well). For favourable CICYs we have a basis $\{J_r\}$ of (1,1)-forms on $X$, obtained from hyperplane classes of the embedding projective spaces, such that the K\"ahler forms $J=t^rJ_r$ correspond to positive values, $t^r>0$, of the K\"ahler parameters $t^r$. The above Mori cone condition can then be written as
\begin{eqnarray} \label{C2}
 \int_X\left( c_2(TX)-c_2(V)\right)\wedge\,J_r \geq0,\text{ for all }r\in\{1,\ldots,h^{1,1}(X)\}\; .
\end{eqnarray} 

\subsection{Constraints from Stability}
Demanding an $\mathcal N=1$ supersymmetric vacuum in four dimensions leads to the requirement that the gauge connection on $V$ satisfies the hermitian Yang-Mills equations at zero slope. By the Donaldon-Uhlenbeck-Yau theorem this is possible if and only if $V$ is holomorphic, has vanishing slope and is polystable. 

The slope of a vector bundle $V$ defined as
$$ \mu(V)  = \frac{1}{\mathrm{rk}(V)} \int_X c_1(V)\wedge J\wedge J =  \frac{1}{\mathrm{rk}(V)} \sum_{r,s,t =1}^{h^{1,1}(X)} d_{rst}\, c_1^r(V) t^s t^t\; ,$$
where  $\displaystyle d_{rst}=\int_X J_r\wedge J_s\wedge J_t$ are the triple intersections on $X$.  \\[-8pt]

For the case of interest, $V$ is a direct sum of line bundles and $c_1(V)=0$. The vanishing slope condition~$\mu(V)=0$ is therefore automatically satisfied. In addition, these sums of line bundles are automatically holomorphic. On the other hand, poly-stability reduces to the requirement that,
\begin{eqnarray} \label{C3}
\exists \, t^r \;\;\; \textnormal{such that} \;\;\; \mu(L_a)|_{t^r}=0  \;\; \forall a
\end{eqnarray} 
somewhere in the interior of the K\"ahler cone ($t^r>0 \;\; \forall r$). 
\vspace{10pt}

Finally, we note that for slope(poly)-stable bundles on a Calabi-Yau threefold there is a positivity condition on the second Chern class, given by the so-called Bogomolov bound \cite{huybrechts2010geometry}. For $SU(n)$ bundles this takes the simple form
\beq\label{bogomolov}
\int_{X} c_2(V) \wedge J \geq 0
\eeq
and $J$ is any K\"ahler form for which $V$ is poly-stable.

\subsection{Constraints from the GUT Spectrum}
The $SU(5)\times S\left(U(1)^5\right)$ GUT spectrum has already been discussed in section \ref{spec}. In order to secure a chiral asymmetry of 3 after taking the quotient of $X$ by $\Gamma$, we must require that $h^1(X,V)-h^2(X,V)=h^1(X,\wedge^2V)-h^2(X,\wedge^2V)=3|\Gamma|$. Since for a poly-stable bundle the zeroth and the top cohomologies vanish, the chiral asymmetry conditions can be formulated in terms of the indices
\begin{eqnarray} \label{C4}
\text{ind}(V) = \text{ind}(\wedge^2V) = -3|\Gamma| \; .
\end{eqnarray}

In fact, for an $SU(5)$ bundle, $\text{ind}(V) = \text{ind}(\wedge^2V)$, so one needs to check only one chiral asymmetry. Furthermore, in order to exclude anti-families, we require the absence of $\overline{\mathbf{10}}$ multiplets, and hence that
\begin{eqnarray} \label{C5}
h^2(X,V)=0
\end{eqnarray}

Finally, in order to safeguard the presence of at least one Higgs doublet, it is necessary to demand the existence of at least one $\mathbf{5}-\mathbf{\overline{5}}$ pair which is expressed by the requirement
\begin{eqnarray} \label{C6}
h^2(X,\wedge^2 V) >0 \; .
\end{eqnarray}

\subsection{Equivariance and the Doublet-Triplet Splitting Problem}

In addition to the above constraints, we demand that, for each $a<b$
\begin{eqnarray} \label{C7}
\text{ind}(L_a\otimes L_b) \leq 0
\end{eqnarray}

In the case where each of the line bundles composing $V$ are individually equivariant, this condition is necessary for it to be possible to project out all Higgs triplets upon the addition of a Wilson line. If the line bundles composing $V$ are individually equivariant, then $V$ descends to a simple sum of line bundles on the quotient space $\hat{X}$. In such a case there is a relation between the index of the line bundle products on $X$ and $\hat{X}$.

$$\text{ind}(\widetilde L_a\otimes \widetilde L_b) = \frac{1}{|\Gamma|}\, \text{ind}(L_a\otimes L_b)$$ 

All indices involved must, of course, be integers and in the case where the line bundles are individually equivariant the size of $\text{ind}(L_a\otimes L_b)$ will be such as to ensure that this is true. Given this relationship, if~$\text{ind}(L_a\otimes L_b)>0$ then so is $\text{ind}(\widetilde L_a\otimes \widetilde L_b)$. This ensures that in such cases there is at least one complete set of ${\bf 5}$ degrees of freedom of $SU(5)$ in the four dimensional effective theory - leading to the presence of Higgs triplets. This result is unaffected by the presence of Wilson lines as the undesirable particle content is protected by an index which such gauge configurations do not affect.

For line bundle sums with non-trivial equivariant blocks the situation is more complicated since divisibility of the index only applies to each equivariant block rather than to individual line bundles. In the simplest such case, an equivariant block is formed by two or more same line bundles which are permuted by the equivariant structure. More complicated equivariant blocks can consist of different line bundles which are mapped into each other, typically subject to an additional permutation of their integer entries. At any rate, bundle isomorphisms between the relevant line must exist in this case so that they must have the same index. In conclusion, line bundles within equivariant block must have the same index and, hence, if this index is positive so is the index of the equivariant block. Then, a generalization of the above index argument to the entire block leads to the same conclusion, namely the inevitable presence of Higgs triplets. Hence, the condition~\eqref{C7} should be imposed in all cases.

\section{The Scanning Algorithm}
\label{sec:alg}
If a sum of five line bundles passes the criteria (\ref{C1} - \ref{C7}), it leads to a consistent four dimensional GUT theory which, with appropriate Wilson line breaking, will lead to heterotic standard models. As mentioned above, a sum of five line bundles is specified by $5\cdot h^{1,1}(X)$ integers. For the manifolds in the CICY database which are favourable and admit known linear free actions of discrete groups, the values of the Hodge number $h^{1,1}(X)$ are restricted~\footnote{Manifolds with $h^{1,1}(X)=1$ cannot lead to consistent line bundle models since the slope zero condition~\eqref{C3} cannot be satisfied.} to the range $2\leq h^{1,1}(X)\leq 6$. Thus, we are interested in investigating bundles described by matrices of between $10$ and $30$ integers, and deciding when they obey the criteria we have described.

One could envisage a scan over all line bundle sums with entries between, say, $-10$ and $10$, for the manifolds with $h^{1,1}(X)=6$. This would require us to check $\sim\!\!10^{30}$ matrices representing sums of line bundles. For comparison, a year has $\sim\!3\cdot 10^{7}$ seconds. It is rather clear that such an attempt would be impossible if one desired to explicitly construct each such line bundle matrix and then check the necessary criteria. A better approach is based on the observation that the criteria (\ref{C1} - \ref{C7}) impose certain conditions on individual line bundles, as well as on partial sums of line bundles. These restrictions are of four kinds: stability related, index related, conditions stemming from the integrability of the heterotic Bianchi identity and restrictions on cohomology. 

\vspace{8pt}
The constraint from stability imposes that each collection of up to five line bundles $\{L_a\}$ can only describe a heterotic vacuum if there exists a point in the interior of the K\"ahler cone, $t^r>0$, such that simultaneously for all $a$, 
$$ \sum_{r,s,t =1}^{h^{1,1}(X)} d_{rst}\, c_1^{\,r}(L_a)\, t^{\,s} t^{\,t} = 0 .$$
This is condition \eqref{C3} of the previous section.

Deciding whether a quadratic equation in several variables has positive solutions can be a computationally intensive question. Establishing the existence of common positive solutions for a collection of such equations is an even more formidable problem. On the other hand, given that the K\"ahler cone for our manifolds is given by $t^r>0$ for all $r=1,\ldots ,h^{1,1}(X)$, a fairly strong, and obviously necessary, condition for the existence of positive solutions to the slope-zero equation for each line bundle $L_a$ is that the matrix $(M_a)_{st}= d_{rst}\,c_1^r(L_a)$ has both positive and negative entries. For any subset $\{L_{a_1},\ldots,L_{a_n}\}\subset \{L_1,\ldots ,L_5\}$ of our five line bundles to have common positive solutions, it is necessary that any linear combination of the matrices $\{M_{a_1},\ldots,M_{a_n}\}$ has both positive and negative entries. In practice, we consider linear combinations with integer coefficients between $-5$ and $5$. This turns out to be a remarkably effective way of eliminating line bundle sums that are not poly-stable. 

For the line bundle sums that pass this necessary criterion we explicitly find common solutions to the slope-zero equations in the interior of the K\"ahler cone. In a great majority of the cases we are able to find exact solutions, while in the remaining cases (most of which appear for the $h^{1,1}=6$ manifolds), we have to resort to numerical methods.

\vspace{8pt}
The index-related criteria impose, for any subset of the five line bundles, that the sum of their indices is negative and greater or equal than $-3|\Gamma|$. This criterion simply follows from equation \eqref{C4} - we do not want more than three generations of standard model particles. We must also check the index based criteria given by equation \eqref{C7}. Indices of line bundles can be computed very rapidly in terms of their defining integers using the following standard formula:
\begin{align*}
\text{ind}(L) & = \frac{1}{12} \left( 2\,c_1(L)^3 + c_1(L)\,c_2(TX)\right)\\
& = \sum_{r,s,t =1}^{h^{1,1}(X)} d_{rst}  \left( \frac{1}{6} \,c_1^{\,r}(L)\,c_1^{\,s}(L)\,c_1^{\,t}(L) + \frac{1}{12} c_1^{\,r}(L)\,c_2^{\,s\,t}(TX) \right)
\end{align*}

\vspace{8pt}
The condition \eqref{C2} stemming from the integrability of the heterotic Bianchi Identity constrains the full sum of five line bundles. This can be rewritten in terms of the integers describing the line bundles as follows.
$$\int_X c_2(TX)\wedge J_r \geq \int_X c_2(V)\wedge J_r = \frac{1}{2}\,d_{rst}\sum_{a=1}^5 c_1^{\,s}(L_a)\, c_1^{\,t}(L_a)$$

The restrictions on the cohomology of the line bundle sums (\ref{C5} - \ref{C6}) require a larger amount of computational resources to implement than do the simple checks already described. Therefore, at first, we only take into account the remaining constraints (\ref{C1} - \ref{C4}) and (\ref{C7}). This stage of the scan leads to GUT models with the correct chiral asymmetry, but does not exclude the possibility of having $\overline{\mathbf{10}}$ multiplets or no Higgs doublets at all. In the second stage of the scan, we attempt to eliminate the models containing anti-families. In $94\%$ of the cases we are able to compute the required cohomology and thus decide upon the fate of the corresponding models. Finally, we eliminate the models that have no Higgs doublets, with a rate of decidability of $88\%$. The computation of the cohomology of line bundles over CICYs is reviewed, for example, in Ref.~\cite{Anderson:2013qca}.

\vspace{8pt}
Below, we schematically present the algorithm used in this automated search. The input parameters are the Calabi-Yau data (configuration matrix, intersection numbers, $c_2(TX)$, a list of row permutations that leave the configuration matrix unchanged); the order of a freely acting discrete group $\Gamma$ and the maximal value for a line bundle entry, $k_{\text{max}}$. The list of permutations present in the Calabi-Yau data is used in order to eliminate redundant line bundle sums, that is, line bundle sums that can be related to one another by a trivial re-labeling of the ambient space projective factors. The algorithm outputs a list of {\tt \itshape Models} represented as matrices of integers whose columns stand for the first Chern classes of the $5$ line bundles.
\vspace{10pt}

{\tt
\begin{itemize}
\item[1.]assemble {\itshape List\_1} containing line bundles satisfying:
\begin{itemize}
\item[$i)$] $-3|\Gamma| \leq \text{ind}\,(L)\leq0$ and 
\item[$ii)$]$\mu(L)=0$, somewhere in the interior of the K\"ahler cone\\[-14pt]
 \end{itemize}

\item[2.] obtain {\itshape List\_1r}$\,\subset\,${\itshape List\_1} by removing all redundant line bundles;

\item[3.]for each $L_{i_1}\in \text{\itshape List\_1r}$ assemble $\text{\itshape List\_2}\,(L_{i_1}) \subset \text{\itshape List\_1}$ containing line bundles such that, for every $L_{i_2}\in \text{\itshape List\_2}\,(L_{i_1})$ the following relations hold: 
\begin{itemize}
\item[$i)$] $-3|\Gamma| \leq \text{ind}\,(L_{i_1})+\text{ind}\,(L_{i_2})$;
\item[$ii)$] $-3|\Gamma| \leq \text{ind}\left(\wedge^2\left(L_{i_1}\oplus L_{i_2}\right)\right) = \text{ind}\,(L_{i_1}\otimes L_{i_2})\leq 0$;
\item[$iii)$]$\mu(L_{i_1})=\mu(L_{i_2})=0$, somewhere in the interior of the K\"ahler cone\\[-14pt]
 \end{itemize}

\item[4.]given $L_{i_1}\in \text{\itshape List\_1r}$, for each $L_{i_2}\in \text{\itshape List\_2}\,(L_{i_1})$ assemble $\text{\itshape List\_3}\,(L_{i_1},L_{i_2}) \subset \text{\itshape List\_2}\,(L_{i_1})$,
such that any  $L_{i_3}\in \text{\itshape List\_3}\,(L_{i_1},L_{i_2})$ satisfies:
\begin{itemize}
\item[$i)$] $-2\,k_{\text{max}} \leq c_1^{\,r}(L_{i_1})+c_1^{\,r}(L_{i_2})+c_1^{\,r}(L_{i_3})\leq2\,k_{\text{max}}$, for all $ r\in\{1,\ldots,h^{1,1}(X)\}$
\item[$ii)$] $-3|\Gamma| \leq \text{ind}\,(L_{i_1})+\text{ind}\,(L_{i_2})+\text{ind}\,(L_{i_3})$;
\item[$iii)$] $-3|\Gamma| \leq \text{ind}(L_{i_1}\otimes L_{i_3})\leq 0$; $-3|\Gamma| \leq \text{ind}(L_{i_2}\otimes L_{i_3})\leq 0$;
\item[$iv)$] $-3|\Gamma| \leq \text{ind}\left(\wedge^2\left(L_{i_1}\oplus L_{i_2}\oplus L_{i_3}\right)\right) = \text{ind}\,(L_{i_1}\otimes L_{i_2})+\text{ind}\,(L_{i_1}\otimes L_{i_3})+\text{ind}(L_{i_2}\otimes L_{i_3})\leq 0$;
\item[$v)$]$\mu(L_{i_1})=\mu(L_{i_2})=\mu(L_{i_3})=0$, somewhere in the interior of the K\"ahler cone\\[-14pt]
 \end{itemize}
 
 \item[5.]given $L_{i_1},L_{i_2}$ and $L_{i_3}$ as above, select from $\text{\itshape List\_3}\,(L_{i_1},L_{i_2})$ those line bundles $L_{i_4}$, such that the line bundle $L_{i_5}$ defined by $c_1(L_{i_1}\oplus L_{i_2}\oplus L_{i_3}\oplus L_{i_4}\oplus L_{i_5})=0$ satisfies: 
\begin{itemize}
\item[$i)$] $-k_{\text{max}} \leq c_1^{\,r}(L_{i_5})\leq k_{\text{max}}$
\item[$ii)$] $-3|\Gamma| \leq \text{ind}\,(L_{i_5})\leq 0$;
\item[$iii)$]$\mu(L_{i_5})=0$ somewhere in the interior of the K\"ahler cone\\[-14pt]
 \end{itemize}
 
 \item[6.] given $L_{i_1},L_{i_2},L_{i_3},L_{i_4}$ and $L_{i_5}$ as above, check: 
 \begin{itemize}
\item[$i)$] $-3|\Gamma| = \text{ind}\,(L_{i_1})+\text{ind}\,(L_{i_2})+\text{ind}\,(L_{i_3})+\text{ind}\,(L_{i_4})+\text{ind}\,(L_{i_5})$;
\item[$ii)$] $-3|\Gamma| = \text{ind}\,\left(\wedge^2 \left( L_{i_1}\oplus L_{i_2}\oplus L_{i_3}\oplus L_{i_4}\oplus L_{i_5} \right) \right)$;
\item[$iii)$] $\text{ind}(L_{i_a}\otimes L_{i_b})\leq 0$ for all pairs $a<b$ that have not been checked so far;
\item[$iv)$]$\mu(L_{i_1})\!=\!\mu(L_{i_2})\!=\!\mu(L_{i_3})\!=\!\mu(L_{i_4})\!=\!\mu(L_{i_5})\!\!\!\!\
=\!0$,  in the interior of the K\"ahler cone\\[-14pt]
\item[$v)$] $\displaystyle 2\,d_{rst}\,c_2^{\,s\,t}(TX) \geq d_{rst}\,\sum_{a=1}^5 c_1^{\,s}(L_{i_a}) \,c_1^{\,t}(L_{i_a})$
 \end{itemize}
 
 if these requirements are satisfied, append $L_{i_1}\oplus L_{i_2}\oplus L_{i_3}\oplus L_{i_4}\oplus L_{i_5}$ to {\itshape Models}
 
 \item[7.] remove all redundant line bundle sums from {\itshape Models}.

 \item[8.] for the remaining {\itshape Models}, check stability by explicitly finding points in the K\"ahler cone where all line bundles have slope zero. 
  
 \item[9.] eliminate models with $\overline{\mathbf{10}}$ multiplets
      
 \item[10.] eliminate models with no $\mathbf{5}-\mathbf{\overline{5}}$ pairs
 
 \end{itemize}
}

\vspace{0.5cm}

In the next section we discuss the results of running this algorithm in addition to the comprehensive nature of the list of models obtained.

\section{Results and Finiteness}\label{sec:results}

It is expected that the number of slope(poly)-stable vector bundles relevant for a smooth heterotic compactification is in fact finite. To begin, it is possible to see that for any bundle which is stable somewhere in the K\"ahler cone, the possible values of its topology, at least, are finite. For example, for the bundles under consideration here, the first Chern class is constrained to vanish (the condition for spinors) and the second Chern class is bounded from above by the anomaly cancellation condition \eref{anom_canc} and from below by the Bogomolov bound \eref{bogomolov}, thus yielding a finite range of possible values for $c_2(V)$. Moreover, for an $SU(n)$ vector bundle which is semistable somewhere in the K\"ahler cone, it is known that there can be only finitely many values possible for the third Chern class \cite{Maruyama,2003math12260L}. Moreover for fixed topology (that is, fixed total Chern class) the moduli space of semi-stable sheaves on a Calabi-Yau threefold is known (by algebraicity of the family \cite{Maruyama,huybrechts2010geometry}) to have only finitely many components. 

An important subtlety arises here for the problem at hand -- namely that the mathematical definition of this moduli space proceeds by first defining an explicit choice of K\"ahler form (that is, a ray in K\"ahler moduli space) with respect to which the sheaves are semistable. However, for the purposes of heterotic model building, we aim to build all Standard Model bundles which are stable for some (not necessarily all the same) K\"ahler form. We wish to know, then, if there are finitely many components to the collection of all moduli spaces, allowing the K\"ahler moduli to vary over {\it any} relevant values in the K\"ahler cone. For this harder problem, some boundedness results are still known (with the most detailed bounds possible in the case of complex surfaces) (see \cite{2003math12260L,huybrechts2010geometry} for a review) for the families of moduli spaces relevant for present scans.

Thus, viewing our poly-stable sums of line bundles as special points in the moduli space of semistable sheaves, it is expected that for a given Calabi-Yau threefold, $X$, there are a bounded number of line bundle Standard Models that can be constructed. However, for the most part, the bounds described above are non-constructive. Thus, in this work, we will empirically bound the number of such models by algorithmic scanning and explicitly constructing the poly-stable vector bundles.

The number of models over a certain manifold admitting discrete symmetries of a fixed order is an increasing and saturating function of the maximal line bundle entry in modulus. This can be observed for all the pairs~$\left( X,|\Gamma|\right)$ that we have considered, as shown in the tables below. However, we believe that our results reflect a more general phenomenon.  In practice, we have applied the above algorithm to all pairs $(X,|\Gamma|)$ of favourable CICYs with the orders $|\Gamma|$ of a freely-acting symmetries and all line bundle sums~\eqref{Vdef} with $|c_1^r(L_a)|\leq k_{\rm m}$, for a fixed upper bound $k_{\rm m}$. In each case, the number of viable models has then been determined for increasing values of $k_{\rm m}$ until saturation occurred. As a practical criterion for the onset of saturation we have required the number of models to remain unchanged for three consecutive values of $k_{\rm m}$. The results, before imposing the absence of $\overline{\bf 10}$ multiplets, Eq.~\eqref{C5}, and the presence of Higgs multiplets, Eq.~\eqref{C6}, can be found in the subsequent tables. The Calabi-Yau manifolds, $X$, are specified by a number, given in the first column of the tables below, which represents their position in the standard list of CICYs compiled in Refs.~\cite{Candelas:1987kf,Green:1987cr} and explicitly accessible here~\cite{Cicydatabase}. As is evident from the tables all viable models consist of line bundles satisfying
\begin{equation}
 |c_1^r(L_a)|\leq 10\; .
\end{equation} 
As one would expect, their number increases dramatically with $h^{1,1}(X)$, the number of K\"ahler parameters. For $h^{1,1}(X)=1,2,3,4,5,6$ we find $0,0,6,552,21731,41036$ models, respectively, for a total of $63325$ models, the number already quoted in the introduction. When the two further constraints~\eqref{C5} and \eqref{C6} are imposed this number reduces to $44343$ and $34989$, as already indicated in Table~\ref{basicstat}. The number of models at each stage, for all pairs $(X,|\Gamma|)$ is tabulated in Appendix~\ref{appC}. The complete list of these models can be accessed here~\cite{lbdatabase}.

We note that the average number of viable models per pair $(X,|\Gamma|)$ as a function of $h^{1,1}(X)$ is approximately given by $0.3,20,530,4560$ for $h^{1,1}=3,4,5,6$, respectively. Very roughly, this corresponding to an increase of one order of magnitude per additional K\"ahler parameter.
At this point it is tempting to speculate about the total number of standard models, that is, models with the MSSM spectrum, in string theory. Known Calabi-Yau three-folds have Hodge numbers in the range of up to $h^{1,1}(X)\leq 500$. If the increase by an order of magnitude observed at small $h^{1,1}(X)$ continues to such large values the number of string standard models is enormous. However, a line bundle sum is determined by $5\cdot h^{1,1}(X)$ integers and it seems likely that the three-family constraint becomes more difficult to satisfy for a large number of these integers. We would, therefore, expect the increase to slow down at larger $h^{1,1}(X)$. Currently, we do not see any way of checking this by explicit scanning since $h^{1,1}(X)=6$ marks out the reach of present computational power. 
{\setstretch{1.35}
\vspace{20pt}
\begin{footnotesize}
\begin{longtable}{| c ||  *{7}{c|}}
\captionsetup{width=14cm}
\caption{\it Number of models as a function of $k_{\text{m}}$ on CICYs with $h^{1,1}(X)=3$. Total number of models: 6}\\
\hline
$\ \ X,\,|\Gamma|\ \ $ & $k_{\text{m}}=1$ & $k_{\text{m}}=2$ & $k_{\text{m}}=3$ & $k_{\text{m}}=4$ & $k_{\text{m}}=5$ & $k_{\text{m}}=6$ & $k_{\text{m}}=7$ \\
\hline
\endfirsthead
\multicolumn{7}{c}%
{\tablename\ \thetable\ -- \textit{Continued from previous page}} \\
\hline
\hline
\endhead
\hline \multicolumn{2}{r}{\textit{Continued on next page}} \\
\endfoot
\hline
\endlastfoot
7484, 4 & 0& 0& 0& 1& 1& 1 & \\ \hline
7669, 3 & 0& 0& 2& 2& 2 & & \\ \hline
7669, 9 & 0& 0& 1& 1& 1 & & \\ \hline
7735, 8 & 0& 0& 0& 0& 1& 1& 1 \\ \hline
7745, 8 & 0& 0& 0& 0& 1& 1& 1 \\ 
\end{longtable}
\end{footnotesize}

\vspace{20pt}
\begin{footnotesize}
\begin{longtable}{| c ||*{9}{c|}}
\captionsetup{width=14cm}
\caption{\it Number of models as a function of $k_{\text{m}}$ on CICYs with $h^{1,1}(X)=4$. Total number of models: 552}\\
\hline
$\ \ X,\,|\Gamma|\ \ $ & $k_{\text{m}}=1$ & $k_{\text{m}}=2$ & $k_{\text{m}}=3$ & $k_{\text{m}}=4$ & $k_{\text{m}}=5$ & $k_{\text{m}}=6$ & $k_{\text{m}}=7$ & $k_{\text{m}}=8$ & $k_{\text{m}}=9$\\
\hline
\endfirsthead
\multicolumn{7}{c}%
{\tablename\ \thetable\ -- \textit{Continued from previous page}} \\
\hline
$\ \ X,\,|\Gamma|\ \ $ & $k_{\text{m}}=1$ & $k_{\text{m}}=2$ & $k_{\text{m}}=3$ & $k_{\text{m}}=4$ & $k_{\text{m}}=5$ & $k_{\text{m}}=6$ & $k_{\text{m}}=7$ & $k_{\text{m}}=8$ & $k_{\text{m}}=9$\\
\hline
\endhead
\hline \multicolumn{2}{r}{\textit{Continued on next page}} \\
\endfoot
\hline
\endlastfoot
6784, 2&0 & 0 & 2 & 10 & 12 & 12 & 12 & &\\ \hline
6784, 4&0 & 6 & 38 & 50 & 62 & 70 & 70 & 70 & \\ \hline
6828, 2&0 & 0 & 1 & 5 & 6 & 6 & 6 & & \\ \hline
6828, 4&0 & 3 & 19 & 25 & 31 & 35 & 35 & 35 &\\ \hline
6831, 2&0 & 1 & 2 & 2 & 2 & & & &\\ \hline
7204, 2&0 & 2 & 14 & 22 & 22 & 22 & & &\\ \hline
7218, 2&0 & 1 & 7 & 11 & 11 & 11 & & &\\ \hline
7241, 2&0 & 1 & 7 & 11 & 11 & 11 & & &\\ \hline
7245, 2&0 & 1 & 4 &  4 & 4 & & & &\\ \hline
7247, 3&0 & 19 & 57 & 59 & 59 & 59 & & &\\ \hline
7270, 2&0 & 2 & 14 &  22 & 22 & 22 & & &\\ \hline
7403, 2&0 & 3 & 6 & 6 & 6 & & & &\\ \hline
7435, 2&0 & 0 & 0 & 2 & 2 & 2 & & &\\ \hline
7435, 4&0 & 0 & 5 & 8 & 9 & 10 & 10 & 10 &\\ \hline
7462, 2&0 & 0 & 0 & 6 & 6 & 6 & & &\\ \hline
7462, 4&0 & 0 & 15 & 24 & 27 & 30 & 30 & 30 &\\ \hline
7468, 2&0 & 5 & 7 & 7 & 7 & & & &\\ \hline
7491, 2&0 & 0 & 0 & 2 & 2 & 2 & & &\\ \hline
7491, 4&0 & 0 & 5 & 8 & 9 & 10 & 10 & 10 &\\ \hline
7522, 2&0 & 0 & 0 & 6 & 6 & 6 & & &\\ \hline
7522, 4&0 & 0 & 15 & 24 & 27 & 30 & 30 & 30 &\\ \hline
7719, 2&0 & 4 & 14 & 26 & 26 & 26 & & &\\ \hline
7736, 2&0 & 2 & 7 & 13 & 13 & 13 & & &\\ \hline
7742, 2&0 & 2 & 7 & 13 & 13 & 13 & & &\\ \hline
7862, 2&0 & 5 & 7 & 10 & 10 & 10 & & &\\ \hline
7862, 4&0 & 9 & 46 & 54 & 58 & 58 & 58 & &\\ \hline
7862, 8&0 & 3 & 40 & 53 & 58 & 62 & 64 & 64 & 64\\ \hline
7862, 16&0 & 0 & 0 & 1 & 4 & 5 & 5 & 5 &\\
\end{longtable}
\end{footnotesize}

\vspace{20pt}
\begin{footnotesize}
\begin{longtable}{| c ||*{9}{c|}}
\captionsetup{width=14cm}
\caption{\it Number of models as a function of $k_{\text{m}}$ on CICYs with $h^{1,1}(X)=5$. Total number of models: 21731}\\
\hline
$\ \ X,\,|\Gamma|\ \ $ & $k_{\text{m}}=1$ & $k_{\text{m}}=2$ & $k_{\text{m}}=3$ & $k_{\text{m}}=4$ & $k_{\text{m}}=5$ & $k_{\text{m}}=6$ & $k_{\text{m}}=7$ & $k_{\text{m}}=8$ & $k_{\text{m}}=9$\\
\hline
\endfirsthead
\multicolumn{7}{c}%
{\tablename\ \thetable\ -- \textit{Continued from previous page}} \\
\hline
$\ \ X,\,|\Gamma|\ \ $ & $k_{\text{m}}=1$ & $k_{\text{m}}=2$ & $k_{\text{m}}=3$ & $k_{\text{m}}=4$ & $k_{\text{m}}=5$ & $k_{\text{m}}=6$ & $k_{\text{m}}=7$ & $k_{\text{m}}=8$ & $k_{\text{m}}=9$\\
\hline
\endhead
\hline \multicolumn{2}{r}{\textit{Continued on next page}} \\
\endfoot
\hline
\endlastfoot
5256, 2&0 & 575 & 727 & 775 & 779 & 779 & 779 & &\\ \hline
5256, 4&0 & 672 & 1857 & 2085 & 2173 & 2180 & 2180 & 2180 &\\ \hline
5301, 2&0 & 144 & 182 & 194 & 195 & 195 & 195 & &\\ \hline
5301, 4&0 & 169 & 466 & 523 & 545 & 547 & 547 & 547 &\\ \hline
5452, 2&0 & 574 & 726 & 774 & 778 & 778 & 778 & &\\ \hline
5452, 4&0 & 672 & 1854 & 2083 & 2171 & 2177 & 2177 & 2177 &\\ \hline
6024, 3&0 & 303 & 510 & 513 & 513 & 513 & & &\\ \hline
6204, 2&0 & 62 & 116 & 122 & 125 & 125 & 125 & &\\ \hline
6225, 2&0 & 147 & 221 & 231 & 232 & 232 & 232 & &\\ \hline
6715, 2&0 & 96 & 148 & 184 & 184 & 184 & & &\\ \hline
6715, 4&0 & 165 & 690 & 812 & 844 & 848 & 848 & 848 &\\ \hline
6724, 2&0 & 19 & 34 & 36 & 39 & 39 & 39 & &\\ \hline
6732, 2&0 & 434 & 778 & 880 & 880 & 880 & & &\\ \hline
6770, 2&0 & 216 & 307 & 329 & 331 & 331 & 331 & &\\ \hline
6777, 2&0 & 434 & 778 & 880 & 880 & 880 & & &\\ \hline
6788, 2&0 & 96 & 148 & 184 & 184 & 184 & & &\\ \hline
6788, 4&0 & 165 & 690 & 812 & 844 & 848 & 848 & 848 &\\ \hline
6802, 2&0 & 432 & 775 & 877 & 877 & 877 & & &\\ \hline
6804, 2&0 & 59 & 154 & 169 & 173 & 173 & 173 & &\\ \hline
6834, 2&0 & 218 & 390 & 441 & 441 & 441 & & &\\ \hline
6836, 2&0 & 24 & 37 & 46 & 46 & 46 & & &\\ \hline
6836, 4&0 & 43 & 175 & 206 & 214 & 215 & 215 & 215 &\\ \hline
6836, 8&0 & 6 & 94 & 120 & 131 & 133 & 137 & 137 & 137\\ \hline
6836, 16&0 & 0 & 0 & 0 & 2 & 3 & 3 & 3 &\\ \hline
6890, 2&0 & 860 & 1546 & 1750 & 1750 & 1750 & & &\\ \hline
6896, 2&0 & 218 & 390 & 441 & 441 & 441 & & &\\ \hline
6927, 2&0 & 144 & 222 & 276 & 276 & 276 & & &\\ \hline
6927, 4&0 & 244 & 1030 & 1212 & 1260 & 1266 & 1266 & 1266 &\\ \hline
6927, 8&0 & 34 & 554 & 706 & 770 & 782 & 806 & 806 & 806\\ \hline
6947, 2&0 & 24 & 37 & 46 & 46 & 46 & & &\\ \hline
6947, 4&0 & 43 & 175 & 206 & 214 & 215 & 215 & 215 &\\ \hline
6947, 8&0 & 6 & 94 & 120 & 131 & 133 & 137 & 137 & 137\\ \hline
6947, 16&0 & 0 & 0 & 0 & 2 & 3 & 3 & 3 &\\ \hline
7279, 2&0 & 128 & 204 & 212 & 218 & 218 & 218 & &\\ \hline
7447, 2&0 & 56 & 87 & 93 & 93 & 93 & & &\\ \hline
7447, 4&0 & 214 & 377 & 419 & 428 & 430 & 432 & 432 & 432\\ \hline
7447, 10&0 & 6 & 58 & 72 & 81 & 82 & 83 & 83 & 83\\ \hline
7487, 2&0 & 277 & 430 & 459 & 459 & 459 & & &\\ \hline
7487, 4&0 & 1052 & 1851 & 2058 & 2101 & 2111 & 2121 & 2121 & 2121\\
\end{longtable}
\end{footnotesize}
}

{\setstretch{1.7}
\vspace{40pt}
\begin{scriptsize}
\begin{longtable}{| c ||*{10}{c|}}
\captionsetup{width=14cm}
\caption{\it Number of models as a function of $k_{\text{m}}$ on CICYs with $h^{1,1}(X)=6$. Total number of models: 41036}\\
\hline
$\ X,\,|\Gamma|\ $ & $k_{\text{m}}=1$ & $k_{\text{m}}=2$ & $k_{\text{m}}=3$ & $k_{\text{m}}=4$ & $k_{\text{m}}=5$ & $k_{\text{m}}=6$ & $k_{\text{m}}=7$ & $k_{\text{m}}=8$ & $k_{\text{m}}=9$ & \myalign{m{1.4cm}|}{$k_{\text{m}}=10,$ $\ \ \ \ 11,\,12,\,13$}\\
\hline
\endfirsthead
\multicolumn{7}{c}%
{\tablename\ \thetable\ -- \textit{Continued from previous page}} \\
\hline
$\ \ X,\,|\Gamma|\ \ $ & $k_{\text{m}}=1$ & $k_{\text{m}}=2$ & $k_{\text{m}}=3$ & $k_{\text{m}}=4$ & $k_{\text{m}}=5$ & $k_{\text{m}}=6$ & $k_{\text{m}}=7$ & $k_{\text{m}}=8$ & $k_{\text{m}}=9$ & \myalign{m{1.3cm}|}{$k_{\text{m}}\!=\!10,$ $\ \ \ \ 11,\,12$}\\
\hline
\endhead
\hline \multicolumn{2}{r}{\textit{Continued on next page}} \\
\endfoot
\hline
\endlastfoot
\varstr{10pt}{5pt} 3413, 3&0& 2278& 2897& 2906& 2906& 2906 & & & &\\ \hline  
\varstr{10pt}{5pt}4190, 2&11& 766&1175& 1243& 1246& 1247& 1249& 1249& 1249 &\\ \hline  
\varstr{10pt}{5pt}5273, 2&29& 4895& 7149& 7738& 7799& 7810& 7810& 7810 & &\\ \hline  
\varstr{10pt}{5pt}5302, 2&0& 4314& 5978& 6360& 6369& 6369& 6369 & & &\\ \hline  
\varstr{10pt}{5pt}5302, 4&0& 11705& 16988& 17687& 17793& 17838& 17868& 17868& 17868 &\\ \hline  
\varstr{10pt}{5pt}5425, 2&0& 2381& 3083& 3305& 3337& 3337& 3337 & & &\\ \hline  
\varstr{10pt}{5pt}5958, 2&0& 148& 224& 240& 253& 253& 253 & & &\\ \hline  
\varstr{10pt}{5pt}6655, 5&0& 92& 178& 189& 194& 194& 198& 201& 202& 203\\ \hline  
\varstr{10pt}{5pt}6738, 2&1& 2733& 4116& 4346& 4386& 4393& 4399& 4399& 4399 &\\
\end{longtable}
\end{scriptsize}
}

\section{An Example}
For illustration, we would like to present a model from our database, which is accessible here~\cite{lbdatabase}. The example is based on the CICY with number 7447, defined by the configuration matrix and line bundle sum
$$
X~=~~
\cicy{\IP^1 \\   \IP^1\\ \IP^1\\ \IP^1\\ \IP^1}
{ ~\bf{1}& \bf{1} \!\!\!\!\\
  ~\bf{1} & \bf{1}\!\!\!\! & \\
  ~\bf{1} & \bf{1}\!\!\!\! & \\
  ~\bf{1} & \bf{1} \!\!\!\!& \\
  ~\bf{1} & \bf{1}\!\!\!\!}_{-80}^{5,45}\; ,
\hskip0.35in
V~=~~
\cicy{ \\ \\ \\ \\ }
{ -1 & -2 & ~~1 & ~~1 & ~~1\\
~~0 & -2 & -1 & ~~1 & ~~2\\
~~0 & ~~2 & -1 & ~~1 & -2\\
 ~~0 & ~~2 & ~~0 & ~~0 & -2\\
 ~~1 & ~~0 &~~ 0 & -2 & ~~1\\}\; .
$$

According to Ref.~\cite{Braun:2010vc}, the manifold $X$ can be smoothly quotiented by a group of order $4$. The columns of the second matrix correspond Êto the first Chern classes of the five line bundles composing $V$. The dimension $h^\bullet(X,V) = \left(h^0(X,V), h^1(X,V), h^2(X,V), h^3(X,V) \right)$ of the bundle cohomologies for $V$ are explicitly given by
\begin{align*} 
h^\bullet(X,V)\ \, & =\  (0,12,0,0) \\
h^\bullet (X,\wedge^2V)&=\ (0, 15, 3, 0)
\end{align*}

The model has a chiral asymmetry of $12$, which, after quotienting, is reduced to $3$. It contains a number of $\bf{5}-\overline{\bf{5}}$ pairs, which after introducing Wilson lines lead to one (or possibly more than one) pair of Higgs doublets. 

\vspace{10pt}
The above example is interesting as it satisfies the anomaly cancellation condition without the addition of any 5-branes. In this case, 
$$c_2(TX).J_i = c_2(V).J_i = \left(24, 24, 24, 24, 24\right) $$ 

As the ranks of $V$ and $TX$ are the same, and their second Chern classes match, one could study the interesting problem\footnote{This idea was suggested to one of us by S.-T.~Yau in a private communication.} of deforming $V$ to $TX$, which would bring us back to the standard embedding. Our database contains 348 such models which saturate the inequality \eqref{C2}.

\section{Final Comments and Outlook}
In this paper, we have presented the results of a comprehensive scan over heterotic line bundle models on favourable complete intersection Calabi-Yau manifolds (CICYs) with freely-acting symmetries. There are $68$ such manifolds with $h^{1,1}(X)=2,\ldots ,6$ contained in the standard list of CICY three-folds~~\cite{Candelas:1987kf,Green:1987cr} available at~\cite{Cicydatabase}. We have focused on rank five line bundle sums, leading to $SU(5)$ GUT models, and a scan over about $10^{40}$ configurations has produced $63325$ consistent and physically viable such models, available here~\cite{lbdatabase}. Furthermore, we have shown computationally that this exhausts the set of physically viable line bundle models on the aforementioned class of CICYs. More precisely, by a consistent and physically viable model we mean a model with a poly-stable line bundle sum which allows for a global completion and whose chiral asymmetries have the correct values to produce a standard model upon taking the quotient by the freely-acting symmetry and including the Wilson line. When we require, in addition, the absence of $\overline{\bf 10}$ multiplets and the presence of at least one ${\bf 5}$--$\overline{\bf 5}$ pair to account for the Higgs the number of viable models is reduced to about 35000.

The task ahead involves constructing the standard models associated to these GUT models. From prior experience with a smaller data set~\cite{Anderson:2011ns, Anderson:2012yf} we expect this will lead to a larger number of standard models compared to the number of GUT models. A number of technical hurdles have to be overcome in order to complete this task, notably devising and implementing a complete  algorithm for computing (equivariant) line bundle cohomology on CICYs. This work is currently in progress. The resulting models will provide by far the largest data set of standard models in any type of string construction and they will provide a starting point for a systematic study of phenomenological questions beyond the spectrum, such as proton decay, $\mu$-problem and the structure of Yukawa-couplings.

We hope to report on the results of this ongoing work in the near future.

\subsection*{Acknowledgments}
L. A. is supported by the Fundamental Laws Initiative of the Center for the Fundamental Laws of Nature, Harvard University. J.~G. would like to acknowledge support by the NSF-Microsoft grant NSF/CCF-1048082. A.~L.~is supported in part by the EC 6th Framework Programme
MRTN-CT-2004-503369 and EPSRC network grant EP/I02784X/1. He would like to thank the string theory group at LMU Munich for hospitality. The work of EP is supported by a Marie Curie Intra European Fellowship within the 7th European Community Framework Programme and by the Heidelberg Graduate School for Fundamental Physics. A large portion of the scan presented in this paper ran on `Hydra', the computer cluster of the Theoretical Physics sub-department at Oxford University. 

\newpage
\appendix

\section{Bundle Structure Groups}\label{Struc_Group_App}
In this Appendix, we review some useful results regarding principal and vector bundle geometry in heterotic compactifications. In particular, we address the problem of how to determine the structure group, $H \subset E_8$, of a vector bundle without knowing an explicit form for the connection.

\subsection{Principal Bundles vs. Vector Bundles}
In compactifications of the heterotic string, for each $E_8$ factor, the gauge fields over the Calabi-Yau threefold are specified by a {\it principal} $H$-bundle, ${\cal V}$, with $H \subset E_8$. Given an explicit embedding of $H$ into $E_8$, ${\cal V}$ determines a collection of associated vector bundles, $V_{\alpha}$, carrying specific representations of $H$, as determined by the decomposition of the $\mathbf{248}$ representation of $E_8$. For example, if a principal $SU(3)$ bundle, ${\cal V}_{SU(3)}$, is embedded into $E_8$ via the direct product $\left( E_6 \times SU(3)\right)/\mathbb{Z}_3$, then the decomposition of the adjoint representation of $E_8$ yields the following representations, carried by the corresponding trio of vector bundles with appropriate rank (fiber dimension):
\beq
\begin{array}{c|c|c}
\varstr{10pt}{5pt} {\bf 3} & {\bf {\overline 3}} & {\bf 8} \\ \hline
\varstr{12pt}{1pt}$\ $ V_{3} $\ \ $ &$\ \ $ V_{3}^{*}$\ \ $&$\ \ $ {\rm End}_{0}(V_3)$\ $
\end{array}
\eeq
Given the rank $3$ vector bundle, $V_3$, in the fundamental representation of $SU(3)$, we can straightforwardly build those bundles corresponding to the ${\bf {\overline 3}}$ and the ${\bf 8}$ by taking the dual or tensor products.

In practice, however, in building the background geometry for a heterotic compactification we do not explicitly construct the principal bundle ${\cal V}$, but rather, first, the vector bundle in the fundamental representation, and from it the full collection of vector bundles, $V_{\alpha}$, in the relevant representations. Moreover, as an added difficulty, except in very special cases, there are no tools available to explicitly construct the $H$-valued connections, $\nabla_{\alpha}$ of the relevant vector bundles\footnote{For numeric approaches to this problem see \cite{Douglas:2006hz,Anderson:2010ke,Anderson:2011ed}.}.

Instead, our starting point is an explicit formal construction of a rank $n$ holomorphic vector bundle (for example a sum of line bundles, or a bundle constructed via a monad \cite{Anderson:2007nc}, or by extension \cite{friedman1998algebraic}). The question now becomes, can we be sure that the given collection of vector bundles really arose from an $H$-valued principle bundle? Suppose, for example, that we consider a holomorphic rank $3$ vector bundle, $V_3$, with structure group~$H \subset U(3)$ and $c_1(V_3)=0$. We may be tempted to declare this an $SU(3)$ vector bundle from this data alone. However, suppose further that the bundle satisfies the non-trivial condition that
\beq\label{self_dual}
V \simeq V^*
\eeq
Now, from this new information, it is clear that the previous conclusion was too hasty. Since the ${\bf 3}$ of $SU(3)$ is not a real representation, it follows that no $SU(3)$ vector bundle can satisfy the self-duality condition in \eref{self_dual}. Instead, the given $V_3$ could actually be carrying the symmetric, ${\bf 3}$-representation of $SU(2)$ (more precisely, it could correspond to $S^2V_2$ for some fundamental, rank $2$, $SU(2)$-bundle, $V_2$); or similarly, the ${\bf 3}$ of an $SO(3)$-bundle. An obstruction of this type could occur for any vector bundle in the collection $V_{\alpha}$, and we must make sure that no such topological obstacles exist in building a bundle with the desired structure group.

\vspace{10pt}
In this work, we focus on $SU(5)$ principal bundles breaking $E_8$ to an $SU(5)$ GUT symmetry in $4$-dimensions via,
\beq\label{248tosu5}
{\bf 248}_{E_8} \rightarrow [({\bf 1}, {\bf 24}) \oplus ({\bf 5}, {\bf \overline{10}}) \oplus ({\bf {\overline 5}}, {\bf 10}) \oplus (\bf{10}, {\bf 5})\oplus ({\bf {\overline{10}}}, {\bf {\overline 5}}) \oplus ({\bf 24},{\bf 1})]_{SU(5)\times SU(5)}
\eeq
Thus, we must construct the associated vector bundles with fiber-dimensions corresponding the ${\bf 5}, {\bf {\overline 5}},{\bf 10}, {\bf \overline{10}}, {\bf 24}$ representations (see Table \ref{spectrum}).

Beginning with the fundamental ${\bf 5}$-representation, for the vector bundles constructed in this work, we will check here that there are no obstructions, such as the one described above, which would prevent the sum of five line bundles, $\bigoplus_{a} L_a$, from having structure group $S \left(U(1)^{\otimes 5} \right)$.

We will outline in the following paragraphs a set of tools for determining the structure groups of rank $n$ holomorphic vector bundles with structure group $H \subset U(n)$ and $c_1(V)=0$. We will focus on distinguishing the groups $SU(n), Sp(2n)$ and $SO(n)$. The exceptional sub-groups of $E_8$ will not arise in the dimensions of representation in consideration here and we will omit them from this discussion.

\subsection{Chern Classes and Structure Groups}

The first and most important ingredient we have in determining the structure group of a vector bundle is its topology. As a simple example, consider the following direct sums of two line bundles on a threefold $X$
\beq
\begin{array}{l|c|c|c}
\varstr{10pt}{5pt} $\ $ V & L_1\oplus L_2 & L \oplus L & L \oplus L^{*}  \\ \hline
\varstr{12pt}{1pt} $\ $ H $\ \ $ & $\ \ $ U(1)\times U(1) $\ \ $ &$\ \ $ U(1)\times U(1)~{\rm or}~ U(1) $\ \ $ &$\ \ $  S[U(1)\times U(1)]=U(1) $\ $
\end{array}
\eeq
For the first sum of line bundles, $c_1(L_1)\neq c_1(L_2)$ implies that for all possible connections on this sum, the structure group is $U(1)\times U(1)$. However, for the sum of two identical line bundles with the same first Chern class, $L\oplus L$, there is some flexibility in the choice of connection. For generic, independent, $U(1)$-valued connections, the structure group likewise is generic, that is, $U(1)\times U(1)$. For this topology, however, a non-generic choice is also available, and by choosing the two connections $\nabla_1=\nabla_2$, the structure group is simply $U(1)$. Finally, in the last example, the sum of a line bundle and its dual, the only structure group compatible with the reducible connection and vanishing trace condition is the diagonal $U(1) \subset SU(2)$.

For phenomenology we require that the low energy GUT symmetry in $4$-dimensions is $SU(5)$ times possible $U(1)$ factors. So long as the commutant of $H$ is of this form, $SU(5) \times S(U(1)^5) \subset E_8$, the Green-Schwarz Mechanism will guarantee that the $U(1)$ symmetries are generically massive (see \cite{Anderson:2011ns,Anderson:2012yf}). Just as in the case of two line bundles described above, here we must guarantee that the topology of our sum does not force a smaller sub-group than $S \left(U(1)^{\otimes 5} \right)$ in such a way that the commutant contains other non-Abelian factors beyond $SU(5)$. For example, if the sum of $5$ line bundles satisfies
\beq
c_1(L_1)+c_1(L_2)+c_1(L_3)=0~~~,~~~c_1(L_4)+c_1(L_5)=0
\eeq
then structure group is $H=S \left(U(1)^{\otimes 3}\right)\times S\left( U(1)^{\otimes 2}\right) \simeq U(1)^{\otimes 3}$, but its commutant in $E_8$ is $SU(6)\times U(1)^{\otimes 3}$ which would not be suitable for model-building. Thus, in the scans outlined in the main body of the text, we have, in addition to $\sum_a c_1(L_a) =0$, imposed that
\beq\label{no_subgroup}
\sum_{a\in S} c_1(L_{a})\neq  0\;\; \text{for all proper subsets}\;\;  S\subset\{1,\ldots ,5\}\; .
\eeq

Finally, having eliminated the possibility of undesirable sub-groups of $S\left(U(1)^{\otimes 5}\right)$ we must still worry about accidental isomorphisms of the form described in \eref{self_dual}, which could force the structure group to be non-unitary and perhaps even larger than $S\left(U(1)^{\otimes 5}\right)$.

\subsection{Hermitian, Real and Symplectic Fiber Structures}
To guarantee that the fibers of $V$ carry the $SU(5)$ representations given in \eref{248tosu5}, we must check that no other topological obstructions, beyond the Chern class conditions described above, exist which could force a different structure group. To begin, we note that we can distinguish between the classical simple groups:~$SU(n), SO(n)$ and $Sp(2n)$ by determining whether the vector bundles carry more than the standard Hermitian fiber metric (characteristic of $U(n)$ bundles \cite{huybrechts2005complex}), but also symplectic or real fiber structures (see \cite{zbMATH01538887} for a review). For example, an $Sp(2n)$-bundle can be represented by a rank $2n$ holomorphic vector bundle (with trivial determinant) equipped with a skew-symmetric, holomorphic pairing, $V \otimes V \to \mathbb{C}$. The pairing can be viewed as an isomorphism $\varphi: V \to V^*$ which is skew-symmetric and non-degenerate on each fiber. The morphism $\varphi$ is referred to as a ``symplectic fiber structure". The case of an $SO(n)$-bundle is identical for rank $n$ holomorphic bundles, where in this case the morphism $\varphi$ is symmetric and forms a ``real fiber structure". These conditions are summarized in \eref{fiber_struc}.
\beq\label{fiber_struc}
\begin{array}{c|c|c|c}
\varstr{10pt}{7pt}  $\ $ Sp(2n)$\ \ $ &$\ \ $  \varphi: V \hookrightarrow \hspace{-8pt} \rightarrow V^* $\ \ $ & $\ \ $ \varphi^*=-\varphi $\ \ $  & $\ \ $  \varphi \in H^0(X, \wedge^2 V) \\ \hline
\varstr{12pt}{1pt} $\ $  SO(n) $\ \ $ &  \varphi: V \hookrightarrow \hspace{-8pt} \rightarrow V^* & \varphi^*=\varphi & $\ \ $  \varphi \in H^0(X, S^2 V)
\end{array}
\eeq

\vspace{5pt}

Returning to the case of $V=\bigoplus_{a=1}^{5}L_a$, no exceptional sub-groups of $E_8$ carry ${\bf 5}$-dimensional representations, so to guarantee that $H \subset S\left(U(1)^{\otimes 5}\right)$ we have only to eliminate the possibility that $V_5$ corresponds to the ${\bf 5}$ of either $SO(5)$ or $Sp(4)$. However, from the conditions above in \eref{fiber_struc} it is clear that this is only possible if $V \simeq V^*$. 

Happily, for the sum of $5$ line bundles considered here such an isomorphism is only possible if $L_a=\cO_X$ for at least one $a \in \{1,\ldots 5\}$. This possibility can be explicitly excluded in scans by demanding that for all $a\in\{1,\ldots ,5\}$
\beq\label{c_not_triv}
c_1^r(L_a) \neq 0 ~~~~\text{for at least one value of}~r\; .
\eeq
Thus, for a sum of five holomorphic line bundles, satisfying $\sum_a c_1(L_a)=0$ in which each summand is a non-trivial line bundle, both real and symplectic fiber structures are not possible.

Having eliminated the possibility of an $SO(n)$ or $Sp(n)$ structure group, via the condition \eref{c_not_triv}, and $V \neq V^*$, and with no exceptional group representations of the appropriate dimension, by process of elimination we have determined that the structure group of $V=\bigoplus_{a=1}^{5} L_a$ satisfies $H \subset SU(5)$. Combining this with the condition \eref{no_subgroup} to exclude undesirable subgroups which might lead to non-Abelian commutants in $E_8$, we have a necessary set of conditions to guarantee a $4d$ GUT symmetry of $SU(5) \times U(1)^4$.

\section{Favourable Embeddings}\label{appB}
In Section \ref{Sec3} we noted that the line bundle scan has been carried out over the class of favourable CICYs, that is, CICYs for which the entire second cohomology descends from the ambient space. We would now briefly like to discuss the precise meaning of this property as well as some criteria which can be used to decide whether a given CICY is favourable. We begin with a CICY, $X$, defined in the ambient space ${\cal A}=\bigotimes_{r=1}^m\mathbb{P}^{n_r}$, as the common zero locus of certain polynomials which can be thought of as sections of the line bundle sum ${\cal N}$ on ${\cal A}$. We denote the restriction of ${\cal N}$ to $X$ by $N={\cal N}|_X$ and also introduce the bundle $S=\bigoplus_{r=1}^m{\cal O}_X({\bf e}_r)^{\oplus (n_r+1)}$, where ${\bf e}_r$ are the standard unit vectors in $m$ dimensions. 

The tangent bundle $TX$ of the CICY $X$ can be obtained from the two short exact sequences
\begin{equation}
 0\rightarrow TX\rightarrow T{\cal A}|_X\rightarrow N\rightarrow 0\;,\quad
 0\rightarrow {\cal O}_X^{\oplus m}\rightarrow S\rightarrow T{\cal A}|_X\rightarrow 0\; .
\end{equation}
 Noting that $H^{1,1}(X)\cong H^2(X,TX)$ and $H^3(X,TX)\cong H^{0,1}(X)=0$ the two associated long exact sequences lead to the following relations for the second cohomology of X
 \begin{eqnarray}
  H^{1,1}(X)&\cong& {\rm Coker}\left(H^1(X,S)\rightarrow H^1(X,N)\right)\oplus{\rm Ker}\left(H^2(X,T{\cal A}|_X)\rightarrow H^2(X,N)\right)\\
  H^2(X,T{\cal A}|_X)&\cong& H^2(X,S)\oplus {\rm Ker}\left(\mathbb{C}^m\rightarrow H^3(X,S)\right)\; .
\end{eqnarray}  
The part of $H^{1,1}(X)$ which descends from the second ambient space cohomology corresponds to the $\mathbb{C}^m$ term in the second equation. Hence, the precise conditions for the CICY $X$ to be favourable are
\begin{equation}
  {\rm Coker}\left(H^1(X,S)\rightarrow H^1(X,N)\right)=0\;,\quad  H^2(X,S)=0\; . \label{favour}
\end{equation} 
In particular, this means a CICY with $h^{1,1}(X)>m$ or $h^1(X,S)<h^1(X,N)$ or $h^2(X,S)>0$ is not favourable. A sufficient, however slightly too strong, condition for $X$ to be favourable is
\begin{equation}
 h^1(X,N)=h^2(X,S)=0\; ,\label{favourprac}
\end{equation} 
where the first of these conditions guarantees that the Coker in \eqref{favour} vanishes. Eq.~\eqref{favourprac} can be checked relatively easily since it only involves cohomologies of line bundles on $X$ and we, therefore, adopt it as our practical definition of favourability.

\section[Distribution of Models]{The Distribution of Models According to $\left(X,|\Gamma|\right)$}\label{appC}

{\setstretch{1.38}
\begin{footnotesize}
\begin{longtable}{| c || c | c | c | c |}
\captionsetup{width=14cm}
\caption{Number of models on CICYs with $h^{1,1}(X)=3$:}\\
\hline
\myalign{| c||}{$\ \ X,\ |\Gamma|\ \ $} &
\myalign{m{2.2cm}|}{$\ $GUT models} &
\myalign{m{3.5cm}|}{ $\ \ \ $ no $ \overline{\mathbf{10}}$ multiplets$\ \ \ $ }&
\myalign{m{3.5cm}|}{$\ \ \ \ \ \ \ $ no $ \overline{\mathbf{10}}\,$s  and $\ \ \ \ \ \ \ $ at least one $\mathbf{5}-\overline{\mathbf{5}}$ pair}&
\myalign{m{3.5cm}|}{$\ \ \ \ \ \ $ no $ \overline{\mathbf{10}}\,$s  and $\ \ \ \ \ \ $ equivariance check for individual line bundles}\\
\hline
\endfirsthead
\multicolumn{5}{c}%
{\tablename\ \thetable\ -- \textit{Continued from previous page}} \\
\hline
\myalign{| c||}{$\ \ X,\ |\Gamma|\ \ $} &
\myalign{m{2.2cm}|}{$\ $GUT models} &
\myalign{m{3.5cm}|}{ $\ \ \ $ no $ \overline{\mathbf{10}}$ multiplets$\ \ \ $ }&
\myalign{m{3.5cm}|}{$\ \ \ \ \ \ \ $ no $ \overline{\mathbf{10}}\,$s  and $\ \ \ \ \ \ \ $ at least one $\mathbf{5}-\overline{\mathbf{5}}$ pair}&
\myalign{m{3.5cm}|}{$\ \ \ \ \ \ $ no $ \overline{\mathbf{10}}\,$s  and $\ \ \ \ \ \ $ equivariance check for individual line bundles}\\
\hline
\endhead
\hline \multicolumn{2}{r}{\textit{Continued on next page}} \\
\endfoot
\hline
\endlastfoot
7484, 4  & 1 & 1 & 1 & 1\\ \hline
7669, 3 & 2 & 2 & 0 (2) & 2 \\ \hline
7669, 9 & 1 & 1 & 0 (1) & 1 \\ \hline
7735, 8 & 1 & 1& 1 &  0 \\ \hline
7745, 8 & 1 & 1 & 1 & 0 
\end{longtable}
\end{footnotesize}

\vspace{10pt}
\begin{footnotesize}
\begin{longtable}{| c || c | c | c | c |}
\captionsetup{width=14cm}
\caption{Number of models on CICYs with $h^{1,1}(X)=4$:}\\
\hline
\myalign{| c||}{$\ \ X,\ |\Gamma|\ \ $} &
\myalign{m{2.2cm}|}{$\ $GUT models} &
\myalign{m{3.5cm}|}{ $\ \ \ $ no $ \overline{\mathbf{10}}$ multiplets$\ \ \ $ }&
\myalign{m{3.5cm}|}{$\ \ \ \ \ \ \ $ no $ \overline{\mathbf{10}}\,$s  and $\ \ \ \ \ \ \ $ at least one $\mathbf{5}-\overline{\mathbf{5}}$ pair}&
\myalign{m{3.5cm}|}{$\ \ \ \ \ \ $ no $ \overline{\mathbf{10}}\,$s  and $\ \ \ \ \ \ $ equivariance check for individual line bundles}\\
\hline
\endfirsthead
\multicolumn{5}{c}%
{\tablename\ \thetable\ -- \textit{Continued from previous page}} \\[8pt]
\hline
\myalign{| c||}{$\ \ X,\ |\Gamma|\ \ $} &
\myalign{m{2.2cm}|}{$\ $GUT models} &
\myalign{m{3.5cm}|}{ $\ \ \ $ no $ \overline{\mathbf{10}}$ multiplets$\ \ \ $ }&
\myalign{m{3.5cm}|}{$\ \ \ \ \ \ \ $ no $ \overline{\mathbf{10}}\,$s  and $\ \ \ \ \ \ \ $ at least one $\mathbf{5}-\overline{\mathbf{5}}$ pair}&
\myalign{m{3.5cm}|}{$\ \ \ \ \ \ $ no $ \overline{\mathbf{10}}\,$s  and $\ \ \ \ \ \ $ equivariance check for individual line bundles}\\
\hline
\endhead
\hline \multicolumn{2}{r}{\textit{Continued on next page}} \\
\endfoot
\hline
\endlastfoot
6784, 2 & 12 & 10  & 10  & 10 \\ \hline
 6784, 4 & 70 & 59  & 59  & 55 \\ \hline
 6828, 2 & 6 & 6  & 6  & 6 \\ \hline
 6828, 4 & 35 & 33  & 33  & 31 \\ \hline
 6831, 2 & 2 & 2  & 2  & 2 \\ \hline 
 7204, 2 & 22 & 14  & 14  & 14 \\ \hline
 7218, 2 & 11 & 11  & 11  & 11 \\ \hline
 7241, 2 & 11 & 9  & 9  & 9 \\ \hline
7245, 2 & 4 & 4  & 4  & 4 \\ \hline
 7247, 3 & 59 & 42 (14) & 22 (4) & 38 \\ \hline
 7270, 2 & 22 & 18  & 18  & 18 \\ \hline
 7403, 2 & 6 & 4 (2) & 0 (3) & 2 \\ \hline
 7435, 2 & 2 & 2  & 2  & 2 \\ \hline 
 7435, 4 & 10 & 9  & 9  & 7 \\ \hline
 7462, 2 & 6 & 6  & 6  & 6 \\ \hline
 7462, 4 & 30 & 16  & 16  & 14 \\ \hline
 7468, 2 & 7 & 6  & 5  & 6 \\ \hline
 7491, 2 & 2 & 2  & 2  & 2 \\ \hline 
7491, 4 & 10 & 4  & 4  & 4 \\ \hline
 7522, 2 & 6 & 6  & 6  & 6 \\ \hline
 7522, 4 & 30 & 21  & 21  & 17 \\ \hline
 7719, 2 & 26 & 24  & 24  & 24 \\ \hline
 7736, 2 & 13 & 12  & 12  & 12 \\ \hline
 7742, 2 & 13 & 12  & 12  & 12 \\ \hline
 7862, 2 & 10 & 10  & 8  & 10 \\ \hline
 7862, 4 & 58 & 53  & 46  & 44 \\ \hline 
7862, 8 & 64 & 52  & 36  & 10 \\ \hline
 7862, 16 & 5 & 5  & 4  & 0 \\ \hline
\end{longtable}
\end{footnotesize}

\vspace{10pt}
\begin{footnotesize}
\begin{longtable}{| c || c | c | c | c |}
\captionsetup{width=14cm}
\caption{Number of models on CICYs with $h^{1,1}(X)=5$:}\\
\hline
\myalign{| c||}{$\ \ X,\ |\Gamma|\ \ $} &
\myalign{m{2.2cm}|}{$\ $GUT models} &
\myalign{m{3.5cm}|}{ $\ \ \ $ no $ \overline{\mathbf{10}}$ multiplets$\ \ \ $ }&
\myalign{m{3.5cm}|}{$\ \ \ \ \ \ \ $ no $ \overline{\mathbf{10}}\,$s  and $\ \ \ \ \ \ \ $ at least one $\mathbf{5}-\overline{\mathbf{5}}$ pair}&
\myalign{m{3.5cm}|}{$\ \ \ \ \ \ $ no $ \overline{\mathbf{10}}\,$s  and $\ \ \ \ \ \ $ equivariance check for individual line bundles}\\
\hline
\endfirsthead
\multicolumn{5}{c}%
{\tablename\ \thetable\ -- \textit{Continued from previous page}} \\[8pt]
\hline
\myalign{| c||}{$\ \ X,\ |\Gamma|\ \ $} &
\myalign{m{2.2cm}|}{$\ $GUT models} &
\myalign{m{3.5cm}|}{ $\ \ \ $ no $ \overline{\mathbf{10}}$ multiplets$\ \ \ $ }&
\myalign{m{3.5cm}|}{$\ \ \ \ \ \ \ $ no $ \overline{\mathbf{10}}\,$s  and $\ \ \ \ \ \ \ $ at least one $\mathbf{5}-\overline{\mathbf{5}}$ pair}&
\myalign{m{3.5cm}|}{$\ \ \ \ \ \ $ no $ \overline{\mathbf{10}}\,$s  and $\ \ \ \ \ \ $ equivariance check for individual line bundles}\\
\hline
\endhead
\hline \multicolumn{2}{r}{\textit{Continued on next page}} \\
\endfoot
\hline
\endlastfoot
5256, 2 & 763 & 625 (12) & 480 (65) & 625 \\ \hline
 5256, 4 & 2128 & 1812 (23) & 1485 (167) & 1444 \\ \hline
 5301, 2 & 191 & 178 (3) & 87 (40) & 178 \\ \hline
 5301, 4 & 534 & 504 (6) & 323 (82) & 406 \\ \hline
 5452, 2 & 762 & 547 (11) & 497 (25) & 547 \\ \hline
 5452, 4 & 2122 & 1624 (17) & 1518 (71) & 1278 \\ \hline
 6024, 3 & 509 & 244 (69) & 215 (29) & 237 \\ \hline
 6204, 2 & 119 & 96 (14) & 76 (17) & 93 \\ \hline
 6225, 2 & 229 & 137 (21) & 118 (17) & 133 \\ \hline
 6715, 2 & 184 & 170 (0) & 138 (4) & 170 \\ \hline
 6715, 4 & 847 & 711 (4) & 539 (76) & 457 \\ \hline
 6724, 2 & 39 & 32 (7) & 20 (10) & 21 \\ \hline
 6732, 2 & 880 & 667 (6) & 532 (60) & 667 \\ \hline
 6770, 2 & 330 & 271 (0) & 197 (39) & 271 \\ \hline
 6777, 2 & 880 & 587 (6) & 549 (32) & 587 \\ \hline
 6788, 2 & 184 & 155 (0) & 147 (4) & 155 \\ \hline
 6788, 4 & 848 & 621 (4) & 579 (28) & 397 \\ \hline
 6802, 2 & 877 & 786 (6) & 524 (128) & 786 \\ \hline
 6804, 2 & 141 & 108 (4) & 99 (5) & 101 \\ \hline
 6834, 2 & 441 & 371 (3) & 283 (47) & 371 \\ \hline
 6836, 2 & 46 & 37 (0) & 36 (1) & 37 \\ \hline
 6836, 4 & 214 & 151 (1) & 147 (4) & 97 \\ \hline
 6836, 8 & 136 & 109 (0) & 97 (9) & 14 \\ \hline
 6836, 16 & 3 & 3 (0) & 2 (1) & 0 \\ \hline
 6890, 2 & 1750 & 1245 (12) & 1091 (83) & 1245 \\ \hline
 6896, 2 & 441 & 421 (3) & 232 (88) & 421 \\ \hline
 6927, 2 & 276 & 243 (0) & 218 (6) & 243 \\ \hline
 6927, 4 & 1264 & 983 (6) & 856 (67) & 628 \\ \hline
 6927, 8 & 798 & 659 (5) & 510 (79) & 81 \\ \hline
 6947, 2 & 46 & 45 (0) & 30 (1) & 45 \\ \hline
 6947, 4 & 214 & 196 (1) & 105 (35) & 127 \\ \hline
 6947, 8 & 136 & 125 (0) & 44 (19) & 21 \\ \hline
 6947, 16 & 3 & 3 (0) & 2 (1) & 0 \\ \hline
 7279, 2 & 218 & 109 (49) & 96 (10) & 108 \\ \hline
 7447, 2 & 93 & 89 (0) & 45 (15) & 89 \\ \hline
 7447, 4 & 430 & 396 (2) & 182 (77) & 306 \\ \hline
 7447, 5 & 0 & 0 (0) & 0 (0) & 0 \\ \hline
 7447, 10 & 81 & 76 (0) & 12 (19) & 0 \\ \hline
 7447, 20 & 0 & 0 (0) & 0 (0) & 0 \\ \hline
 7487, 2 & 459 & 319 (0) & 261 (28) & 319 \\ \hline
 7487, 4 & 2115 & 1505 (8) & 1257 (94) & 1136 \\
\end{longtable}
\end{footnotesize}

\vspace{10pt}
\begin{footnotesize}
\begin{longtable}{| c || c | c | c | c |}
\captionsetup{width=14cm}
\caption{Number of models on CICYs with $h^{1,1}(X)=6$:}\\
\hline
\myalign{| c||}{$\ \ X,\ |\Gamma|\ \ $} &
\myalign{m{2.2cm}|}{$\ $GUT models} &
\myalign{m{3.5cm}|}{ $\ \ \ $ no $ \overline{\mathbf{10}}$ multiplets$\ \ \ $ }&
\myalign{m{3.5cm}|}{$\ \ \ \ \ \ \ $ no $ \overline{\mathbf{10}}\,$s  and $\ \ \ \ \ \ \ $ at least one $\mathbf{5}-\overline{\mathbf{5}}$ pair}&
\myalign{m{3.5cm}|}{$\ \ \ \ \ \ $ no $ \overline{\mathbf{10}}\,$s  and $\ \ \ \ \ \ $ equivariance check for individual line bundles}\\
\hline
\endfirsthead
\multicolumn{5}{c}%
{\tablename\ \thetable\ -- \textit{Continued from previous page}} \\[8pt]
\hline
\myalign{| c||}{$\ \ X,\ |\Gamma|\ \ $} &
\myalign{m{2.2cm}|}{$\ $GUT models} &
\myalign{m{3.5cm}|}{ $\ \ \ $ no $ \overline{\mathbf{10}}$ multiplets$\ \ \ $ }&
\myalign{m{3.5cm}|}{$\ \ \ \ \ \ \ $ no $ \overline{\mathbf{10}}\,$s  and $\ \ \ \ \ \ \ $ at least one $\mathbf{5}-\overline{\mathbf{5}}$ pair}&
\myalign{m{3.5cm}|}{$\ \ \ \ \ \ $ no $ \overline{\mathbf{10}}\,$s  and $\ \ \ \ \ \ $ equivariance check for individual line bundles}\\
\hline
\endhead
\hline \multicolumn{2}{r}{\textit{Continued on next page}} \\
\endfoot
\hline
\endlastfoot
 3413, 3 & 1737 & 709 (516) & 599 (98) & 698 \\ \hline
 4190, 2 & 1145 & 540 (195) & 473 (57) & 429 \\ \hline
 5273, 2 & 6753 & 4154 (934) & 3292 (701) & 3757 \\ \hline
 5302, 2 & 6294 & 4130 (246) & 3291 (456) & 4130 \\ \hline
 5302, 4 & 17329 & 13242 (82) & 10174 (1678) & 9235 \\ \hline
 5425, 2 & 3128 & 1946 (533) & 1358 (409) & 1802 \\ \hline
 5958, 2 & 246 & 215 (23) & 103 (66) & 179 \\ \hline
 6655, 5 & 161 & 143 (15) & 67 (64) & 1 \\ \hline
 6738, 2 & 4243 & 1846 (743) & 1599 (169) & 1763 \\
\end{longtable}
\end{footnotesize}

}


\newpage

\bibliography{bibfile}{}
\bibliographystyle{utcaps}
\end{document}